# Semantic orchestration and exploitation of material data: A dataspace solution demonstrated on steel and copper applications

*Yoav Nahshon\*, Lukas Morand, Matthias Büschelberger, Dirk Helm\*, Kiran Kumaraswamy, Paul Zierep, Matthias Weber, Pablo de Andrés*

Y. Nahshon\*, L. Morand, M. Büschelberger, D. Helm\*, K. Kumaraswamy, P. Zierep, M. Weber, P. de Andrés
Fraunhofer Institute for Mechanics of Materials IWM, 79108 Freiburg, Germany

Corresponding authors E-mail:

dirk.helm@iwm.fraunhofer.de , yoav.nahshon@iwm.fraunhofer.de

**Abstract**

In materials science and manufacturing, vast amounts of heterogeneous data (e.g., measurement and simulation logs, process data, publications) serve as the bedrock of valuable knowledge for various engineering applications. However, efficiently storing and managing this diverse data poses challenges due to limited standardization and integration across different organizational units. Addressing these challenges is essential to fully unlock the potential of data-driven approaches. This paper introduces novel, comprehensive semantic methodology tailored to materials engineering and realized as a technology stack named Dataspace Management System (DSMS), which powers dataspace solutions that leverage the knowledge encoded in heterogeneous data sources to support data-driven insights and to derive new knowledge. At its core, DSMS offers a distinctive knowledge management approach tuned to meet the specific requirements of the materials science and manufacturing domain, all while adhering to the FAIR principles. DSMS provides functionalities for data integration, linkage, exploration, visualization, processing, data sharing, and services (e.g., consulting) to support engineers in decision-making, design and optimization. We present an architectural overview of DSMS, outlining its core concepts and their technological implementation, as well as demonstrate its applicability to common data-processing tasks through use cases from the StahlDigital and KupferDigital research projects within Germany's MaterialDigital initiative.

# 1. Introduction

## 1.1. Motivation and state-of-the-art

Developing and producing materials and products in a sustainable and resilient manner is a major challenge in materials science and manufacturing, essential for driving the green transformation of the industry while maintaining competitive advantages. Over the last few decades, initiatives such as Integrated Computational Materials Engineering (ICME) [1], which focusses on virtual representations of materials and processes, and the digitalization of manufacturing processes via smart factories [2], often referred to as Industry 4.0/5.0, have made significant strides towards achieving this transformation. However, one of the most significant challenges that has emerged is managing the heterogeneous material and process data originating from various sources, including production process data, data from various repositories, and data relevant during the materials, process, and product design phases, such as experimental measurements, simulation data, logs, and domain knowledge. To overcome this challenge and enable FAIR data (findable, accessible, interoperable, reusable) [3], our research hypothesis proposes that a solution can be achieved by leveraging semantic technologies to implement a novel knowledge management approach for dataspaces. In this context, the term *dataspace* refers to the topics and challenges in the data management field related to data integration and interoperability [4].

In this paper, we present a semantics-based approach supported by the established effectiveness of knowledge graphs in knowledge management [5]. By transforming the typically acquired characterization, modelling, and simulation data from the syntactic raw data level to a semantic level, FAIR data principles can be ensured. Interoperability driven by semantic technologies is especially advantageous, as it enables seamless communication and knowledge retrieval from diverse sources. This approach ultimately facilitates more efficient data analysis and artificial intelligence application [6], fully unlocking the potential within the data. Examples include learning constitutive relations [7], solving materials and process design problems [8, 9], material model parameter identification [10], analyzing microstructure image data [11], compressing microstructure information [12], design space exploration [13, 14], and solving ill-posed inverse problems [15]). While the usage of semantic technologies is still in its early stages in materials science and manufacturing, it already demonstrated success in various domains over the past two decades [15], including bioinformatics [16], multilingual databases like BabelNet [17], and general knowledge bases like Wikidata [18] and DBpedia [19] as well as in enterprises applications [20]. Consequently, many technology companies now employ knowledge graphs to link data and extract valuable insights [21–24].

Semantic technologies are essential for achieving FAIR data management, as they involve the use of common ontologies to define and standardize terms and relationships within the data [25]. An important prerequisite is the establishment of shared ontologies, ensuring users agree on standardized terms to describe both the data itself and the relationships between different data points. Ontologies have gained popularity in materials science in recent years, particularly within the European materials science community, for their ability to standardize data [26]. It is envisioned that data from different sources and data from various sources and locations (e.g., data silos) can be made accessible through semantic enhancement (i.e., via ontologies), facilitating the development of novel applications in materials science and manufacturing [27]. Adopting an ontology-based representation of the data results in a uniform structure that helps reconciling incompatible data formats and schemes. Additionally, ontologies significantly enhance the FAIRness of data by making data content interpretable by systems, thereby achieving machine-readability. This, in turn, offers new opportunities for deeper understanding and insights based on the underlying data.

Applying ontologies to represent data in a semantic-enhanced manner is, nevertheless, not straightforward [28]. Adding this representation layer to the system immediately raises concerns about scalability and maintainability - two key factors for any system operating in a production environment. Experience shows that upper- and middle-level ontologies, such as the Elementary Multiperspective Materials Ontology (EMMO) [29], the Basic Formal Ontology (BFO) [30] and the Platform Material Digital (PMD) core ontology [31], are often too generic to capture all terms relevant to specific experimental or simulation data. To account for these gaps, domain and ontology experts are required to fill in missing details by developing application ontologies, which introduce new and more specific subclasses to the upper- and middle-level ontologies. However, even for data describing similar procedures (e.g., tensile test data originating from two different machines), it is often necessary to adapt or develop application ontologies almost from scratch due to variations in the raw data, the use cases, and the work procedures.

In contrast to semantics-based dataspace systems, database solutions utilizing the traditional relational database model have gained traction in the past years [32]. Furthermore, to standardize experimental data generation, the adoption of electronic lab notebooks (ELN) has increased over the past two decades [33]. While existing industrial and commercial tools for handling data are often mature, they typically do not adhere to FAIR data principles. Consequently, although these tools perform effectively for their specific purposes, they face limitations in terms of interoperability. These solutions are commonly rely on a traditional SQL and NoSQL database, complemented by a graphical and a programming user interfaces, as outlined in a recent study [34]. These systems are primarily designed to accommodate and manage data within an organizational unit (e.g., a company or a use case), offering a rigid structure to which integration workflows must conform.

In the context of the data-information-knowledge-wisdom (DIKW) hierarchy [35], traditional database solutions operate at a relatively low information level. For instance, a data point may have a label and a fixed unit but lack contextual information. In contrast, achieving broader applicability and transferability of knowledge requires operating at higher levels within the DIKW hierarchy, where the meaning of data alongside logical relations are also captured. This holds true also for data sharing, where even at the institutional level, data is often collected using different techniques in non-standardized formats (using CSV and Excel files, hand-written protocols,). Integration, whether on a small within an organization or across company boundaries, remains challenging with traditional, yet mature, database solutions, due to limitations in knowledge sharing and insufficient descriptive metadata.

**1.2. Contribution and advantages of using the Dataspace Management System**

In this paper, we present the Dataspace Management System (DSMS), a novel technology stack that powers dataspace solutions, customizable to meet the specific demands of the relevant domain. DSMS provides a solution to the challenges outlined in our research hypothesis stated above (handling heterogeneous data from various sources and enabling FAIR data representations, etc.) by utilizing semantic technologies and a knowledge graph-based concept as the basis of the dataspace systems. DSMS stands out among existing dataspace solutions due to its unique approach to knowledge management. The approach involves fragmenting knowledge into discrete, self-contained objects that are interconnected, forming a network of resources. This paper demonstrates how complex workflows in materials engineering can be efficiently represented by DSMS. Figure 1 illustrates the key features of the dataspace solution, including data integration and linkage, data exploration and the generation of new knowledge, ultimately supporting wisdom through recommendations and decision support.

The functionalities of DSMS bridge the gap between (traditional) standalone database solutions, electronic lab notebooks (ELN) and initiatives like the International Data Spaces (IDS), where the term *dataspace* (or *data space* according to IDS spelling) takes on a different meaning, emphasizing data sovereignty and access rights [36]. Unlike ELNs, which primarily focus on recording, managing, and sharing experimental data within a laboratory setting, DSMS offers a more comprehensive approach by enabling to manage and integrate data from various sources, support complex workflows, and facilitate knowledge generation and decision support. It is important to note that in the present work, the term dataspace is used to describe a complete system that includes a data (storage) layer, a data processing (business) layer, and a representation layer. In contrast, standard database approaches typically include only a data layer, and larger dataspace initiatives like the International Data Space (IDS) focus more on linking different types of data sources, including capabilities like data brokerage for fine tuning data-sharing policies.

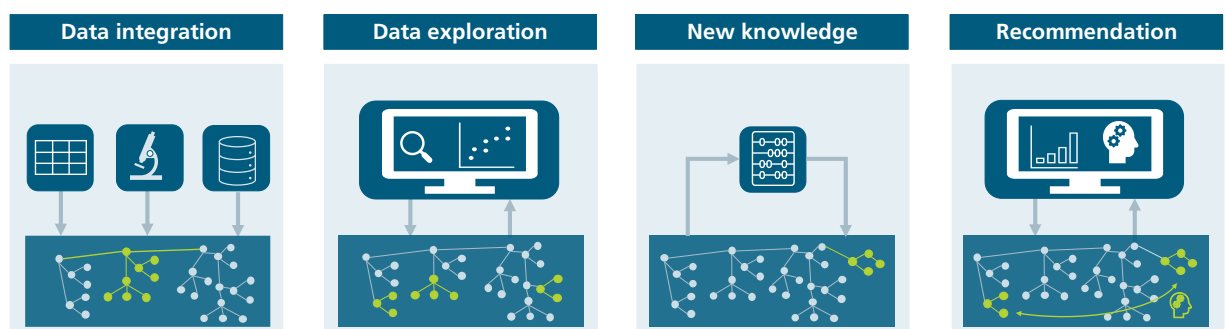

**Figure 1.** Key features of a dataspace solution powered by the Dataspace Management System (DSMS)

### 1.3. Paper structure

This paper is organized as follows: In Section 2, we provide a high-level overview of the concepts and methodology implemented by the Dataspace Management System (DSMS) to power dataspaces. In Section 3, we demonstrate its capabilities through selected use cases involving steel and copper materials from the two research projects StahlDigital [37] and KupferDigital [38], funded by the German MaterialDigital initiative. In Section 4, we discuss the results and give an outlook on the future prospects of DSMS in Section 5.

### 2. Conceptual and methodological framework of Dataspace Management System (DSMS)

In this section, we introduce the Dataspace Management System (DSMS) as a combined reference to both the conceptual methodology and its software implementation. For clarity and convenience, "DSMS" will serve as shorthand for the full methodological framework as realized in software. We begin with a software engineering and development perspective to establish robust design principles that ensure practical usability given the complexity of materials science data. This perspective is essential for effectively meeting requirements for scalability, integration, and flexibility. We then outline the requirements derived from materials science applications and production processes, highlighting the practical needs that DSMS must address to have a meaningful impact in real-world scenarios. Finally, we present the key concepts and components of DSMS, providing an overview of its methodology and its core elements that enable data-driven insights and support decision-making. The usefulness of the concepts and methodology introduced here will be validated later in Section 3 through specific use cases.

## 2.1. Software engineering perspective

The software engineering of DSMS and its derived dataspaces follow clear development procedures that adhere to industry standards, such as ISO/IEC/IEEE 42010 "Software, systems and enterprise - Architecture description" [39]. This standard provides a structured framework for describing the architecture of software systems, ensuring a comprehensive understanding of the system's design and implementation from multiple perspectives, or viewpoints, as they are referred to by the standard. Although our primary focus in this paper is the scientific contribution, highlighting a few key architectural aspects can attest to the solution's maturity and practical applicability, illustrating how it meets the demands of scalability, integration, and flexibility required for materials science data. The following summarizes a few selected key aspects specified by ISO/IEC/IEEE 42010 as they apply to DSMS:

- **Stakeholders:** The architecture of the system must consider a diverse set of stakeholders across both manufacturing companies and research institutes. These stakeholders have varying needs, roles, and expectations from the system, which drive the overall design and functionality. The identified stakeholders are software engineers, materials engineers/scientists/domain experts, data scientists, lab technicians, R&D managers, CEOs.
- **Concerns:** The system must address several critical concerns for the different stakeholders, ensuring that the architecture supports their needs both functionally and technically. The concerns with respect to the identified stakeholders are: scalability, maintainability, data privacy and security, usability, meaningful technical insights, data quality, compliance (e.g., GDPR, DIN norms), business insights, and FAIR data management.
- **Viewpoints:**

- o **Functional viewpoint.** This viewpoint focusses on expected functionalities, including data integration, data retrieval, analysis, post-processing (e.g., generating new knowledge from existing data), and the ability to obtain an overview of the product life cycle, as well as processes within each step (e.g., manufacturing processes).
- o **Information viewpoint.** This viewpoint outlines how data is organized, stored, and accessed within the system. Ontologies and semantic web technologies play an important role in achieving data FAIRness and enabling data traceability, which in turn contributes to quality and reliability of the data.
- o **Deployment viewpoint:** DSMS enables deployments of tailored dataspaces in cloud environments, on-premises, or in hybrid configurations, ensuring flexibility and adaptability to different organizational needs.

In addition to following standards, learning from experience also plays an important role in developing effective systems. While semantic technologies hold potential for many desirable features, they also come with high costs, namely, extensive development periods of ontologies, difficulties in maintaining and adapting them, and a lack of suitable tools [40]. To mitigate these shortcomings, DSMS employs a bottom-up approach for semantic integration of data in an incremental manner. This bottom-up approach prioritizes the data rather than its semantic representation, which is particularly useful for the onboarding of newcomers into the system. In contrast, a top-down approach first focuses on developing the semantic framework that fits the use case in hand, typically by following a common ontology development methodology, such as NeOn [41] and LOT [42]. In recent years, many initiatives within the materials science community followed the top-down philosophy, resulting in overly intensive and complex workflows, making them unadoptable beyond the context of the corresponding projects. Furthermore, a fully pledged semantic representation is not always necessary, or rather required only for a few specific cases.

To develop an effective system following a bottom-up approach it is therefore useful to classify important features of the system, namely by the levels of semantic integration of data. Such classification provides orientation on the supported features per level, allowing stakeholders to efficiently progress through the various levels according to their needs and resources. Table 1 provides a breakdown of the integration levels to efficiently handle the complexity of the enhanced semantic representations.

**Table 1.** Integration levels of the sematic integration of data

| Integration level | Description | Application/Example |
|---|---|---|
| 0 | Data can be uploaded | An uploaded file is accessible in the system |
| 1 | Semantic annotation of the data/ data file type | The system knows that the uploaded file is in a csv format |
| 2 | Semantic annotation of the context of the data/ data file | The system knows that the uploaded file is a tensile test file |
| 3 | Semantic annotation of the metadata | The system knows the metadata of the tensile test file |
| 4 | Semantic annotation of the metadata and accessibility of all data in an interoperable manner | The system knows the metadata of the tensile test file and the time series of the file are accessible |
| 5 | Complete semantic annotation | The system knows all information in the tensile test file: i.e. the meta data and the time series. |

## 2.2. Requirements derived from materials science applications and production processes

The development of DSMS and the dataspaces it powers must address specific requirements derived from its applications and end-users to ensure its effectiveness and relevance in the materials science and production engineering field. These requirements can be categorized into several key areas:

- **Data heterogeneity and integration:** In materials science and production engineering, there is a vast amount of heterogeneous data that is either required or generated. This includes experimental measurements, simulation results, machine data, and publications, which are often stored across various data repositories. Therefore, DSMS must be capable of integrating diverse data types and formats into a unified system. This includes support for various file formats (e.g., CSV, Excel, JSON, XML), various file conventions (e.g., line endings in Linux and Windows) and the ability to semantically annotate and link data from different sources.
- **FAIR data principles:** Ensuring that data adheres to FAIR principles (Findable, Accessible, Interoperable, Reusable) is the key for data sustainability. DSMS must provide robust mechanisms for finding data, access control, interoperability between different data sources, and mechanisms for reusing data. This includes implementing standardized metadata schemas, persistent identifiers, and interoperability frameworks.
- **Scalability and performance:** DSMS must be scalable to handle large volumes of data and high-frequency data generation, which is typical in materials science and production environments. It should offer efficient data storage solutions and high-performance data processing capabilities to ensure smooth and responsive user experiences even with extensive datasets.

- **User-friendly interfaces:** The system must cater to a wide range of users, including experimenters, materials scientists, data scientists, engineers, and product designers. DSMS should provide intuitive and user-friendly interfaces that allow users to store, access, explore, and analyze data with minimal technical expertise. This includes graphical user interfaces (GUIs) and APIs for advanced usage.

- **Automated data processing and analysis:** DSMS should support automated workflows for data processing and analysis. This includes tools for data cleaning, transformation, and enrichment, as well as capabilities for executing complex analytical tasks, such as machine learning models and simulations. Automation reduces the manual effort required for repetitive tasks and ensures consistency and accuracy in data handling.

- **Provenance and traceability:** Keeping track of data provenance is essential for ensuring the reproducibility and reliability of research outcomes. DSMS must provide comprehensive logging and tracking of data manipulations/ transformations, including data sources, processing steps, and user interactions. Generally, traceability is crucial for auditing, troubleshooting, and validating results.

- **Collaboration and data sharing:** Collaboration among different stakeholders is vital for increasing productivity and innovation. DSMS should facilitate seamless data sharing and collaboration by providing access control mechanisms, versioning, and collaborative tools. This enables multiple users to work on shared datasets while maintaining data integrity and security.

- **Security and data sovereignty:** Ensuring data security and respecting data sovereignty are critical factors for organizations when adopting new solutions. DSMS must implement robust security measures to protect sensitive data from unauthorized access and breaches. Additionally, it should comply with legal and regulatory requirements related to data ownership and sovereignty, allowing users to control how and where their data is stored and processed.

Addressing these requirements for achieving an effective solution is, however, not straightforward. Along the value chain from raw material to a finished product, DSMS must accommodate the various user requirements to ensure comprehensive support. Users typically need to integrate different types of data, including material testing results, simulation data and general information such as from material data sheets. Establishing relationships between data points, such as linking tensile test data with testing machines and evaluation procedures, is essential. Users must be able to explore data effectively, for instance, by plotting typical relations between processing parameters or conditions, microstructural features of a material and material or product properties. Moreover, enriching data by calculating secondary parameters, such as the Young's modulus, yield strength or fracture toughness, is crucial for in-depth analyses. Users also wish to get recommendations based on available data from the system, such as the optimal material for a component produced by a certain manufacturing process. Maintaining control over stored data is paramount, ensuring data sovereignty and compliance with company policies and security standards. These user requirements align closely with the FAIR principles, where effective data exploration demands findability and accessibility, and seamless data enrichment necessitates interoperability and reusability. Addressing these needs ensures that DSMS can support users throughout the entire product life cycle, from raw material to the product, enhancing both efficiency and innovation in materials science and production engineering.

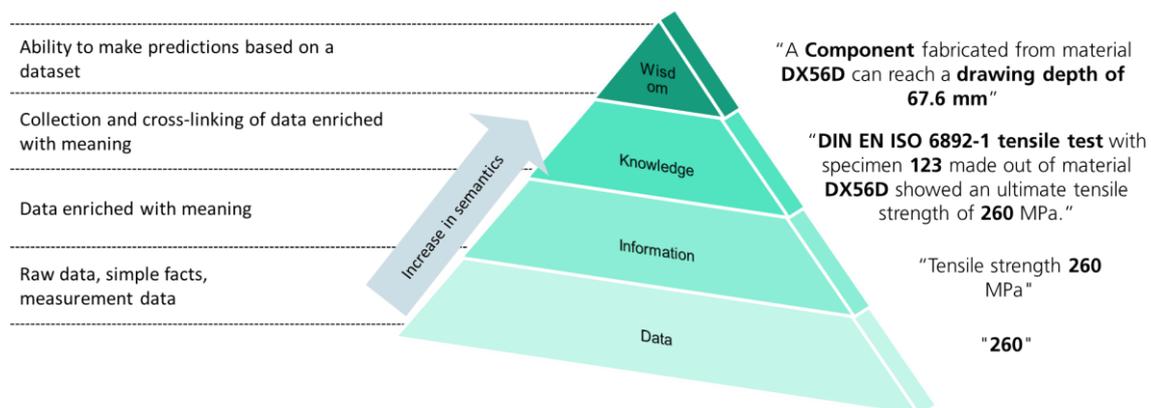

**Figure 2.** Data-information-knowledge-wisdom (DIKW) hierarchy applied to materials data

On a semantics level, the Data-Information-Knowledge-Wisdom (DIKW) [35] hierarchy describes the transformation of raw data into actionable insights and decision-making. This is illustrated in Figure 2 at the example of materials data. DSMS leverages this hierarchy to enhance the value derived from data at each level:

1. **Data:** At the base level, data consists of raw facts collected from various sources such as experimental measurements, simulation outputs, and sensor readings. In materials science, this could include test results, compositional analysis, and microstructural images etc. DSMS ensures that this data is collected, stored, and made accessible in a standardized format.

2. **Information:** When data is processed and organized, it becomes information. For instance, raw tensile test data is enriched with meaning by giving it a label and a unit. DSMS facilitates the transformation of raw data into information through automated processing and data integration tools, making it easier for users to interpret and analyze the data.
3. **Knowledge:** Information gains context and becomes knowledge when it is analyzed and linked to existing theories, models, and datasets. For example, gaining knowledge on how mechanical properties of a material relate to a testing process, as well as understanding how these properties relate to microstructure and composition, involve synthesizing information from multiple sources. DSMS supports knowledge generation by enabling the creation of semantic links between other resources, such as materials models, allowing users to draw connections and derive deeper insights.
4. **Wisdom:** At the highest level, wisdom refers to making informed decisions and predictions based on the accumulated knowledge. In production engineering, this means, for example, to select the optimal material for a specific application based on its performance characteristics (and cost considerations). DSMS provides tools for advanced data analysis, machine learning, and simulation, empowering users to make data-driven decisions and optimize production processes.

## 2.3. Core concepts and methodological design of the Dataspace Management System

The main goal of DSMS as a technology stack is to power comprehensive dataspaces for managing complex data workflows in materials science and manufacturing. To achieve this, DSMS incorporates several key concepts and components that collectively ensure efficient data integration, processing, visualization, and knowledge generation. This subsection outlines these core concepts and components.

### 2.3.1. Knowledge management concepts

At the heart of DSMS's knowledge management philosophy lie two central definitions: *knowledge type* and *knowledge item*, respectively referred to as *k-type* and *k-item* for convenience. The conceptual element *k-type* is used to classify materials information. Example of such classes include chemical element, material, process, and material test. The k-types are determined by domain requirements to structure knowledge in a meaningful manner. K-types also provide a specification template to represent objects accurately within the system. For example, the k-type "material" has an attribute "composition", while the k-type "organization" may include a (geographical) "location" as a data attribute. While k-types provide the template, *k-items* represent instances of information that follow the template of their corresponding k-types. In other words, each k-item is associated to a particular k-type, containing data aligned with its k-type specifications. K-items are designed to encapsulate data, metadata, and semantic information, providing a comprehensive and flexible means to manage diverse pieces of knowledge.

Each k-item consists of three main parts: metadata, data container, and semantic graph, as depicted in Figure 3. The following is a breakdown of these components.

- **Metadata:** The metadata component of a k-item serves as the descriptive layer, providing essential information about the data it contains. In DSMS, such information is stored in an SQL database, providing a robust and highly efficient data layer. Key elements of the metadata include:
    - Identifier: A unique identifier for the k-item, ensuring that it can be distinctly referenced within the dataspace.
    - K-type: The classification of the k-item, such as an organization, expert, or dataset, allowing for structured categorization. Knowledge item types are also referred to as k-types for short.
    - Summary: A brief description of the k-item, summarizing its content and purpose for quick understanding.
    - Links to other k-items: Relationships to other k-items that may have a label and a corresponding ontology equivalent. Together they establish a network of interlinked knowledge within the dataspace.
    - Associated apps: Applications that are relevant to the k-item, facilitating direct interaction and processing of the contained data.
    - Semantic annotations: Annotations that provide semantic context to the k-item, enhancing its discoverability and integration within the dataspace.
- **Data container:** The data container is the core storage component of a k-item, capable of holding files in various formats. This flexibility ensures that diverse types of data can be accommodated. Key features of the data container include:
    - Files: The data container can store files in any format, including text, images, videos, and specialized scientific data formats.
    - Build-in support for multidimensional array data: For effectively handle complex, multidimensional array data, such as time series of a tensile test data

including stress-strain measurements, enabling AI-ready data processing and seamless analysis.
- **Semantic graph:** The semantic graph component represents the data within the k-item in a structured, machine-readable format. This facilitates advanced data integration, retrieval, and analysis. Key features of the semantic graph include:
    - An RDF representation: The data is represented using the Resource Description Framework (RDF), a standard model for data interchange on the web. This ensures interoperability and ease of integration with other semantic web technologies. This representation is curated by a dedicated triplestore database, shipped with DSMS, to enable direct queries against it using the standard SPARQL query language. The representation may also be stored as RDF files (e.g., in turtle format) in the data container of the k-item for documentation and specific applications.
    - References to data points: The semantic graph can reference specific data points within the data container or external data sources, allowing for precise linkage and contextualization. This capability is important for efficiency as it allows the RDF file to remain relatively small while maintaining links to large data chunks such as time series data.
    - Adherence to common vocabularies and ontologies: The semantic graph adheres to established vocabularies and ontologies, ensuring consistency and enhancing the semantic richness of the data.

To elucidate the concept of k-items, consider the example given in Table 2. Example k-item for a dataset originating from a tensile test.

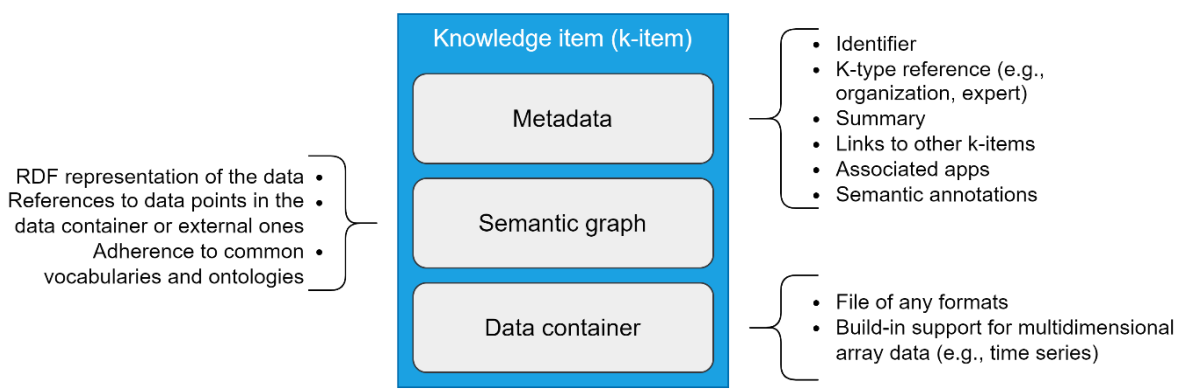

**Figure 3.** The internal structure of knowledge items (k-items). Each k-item consists of three parts: (i) metadata, (ii) semantic graph, and (iii) data container. The metadata includes a reference to a corresponding knoweldge type (k-type).

**Table 2.** Example k-item for a dataset originating from a tensile test.

| Metadata | - **Identifier:** Dataset-001<br>- **K-type:** Dataset<br>- **Summary:** Tensile test data for sample XYZ produced by Kappa machine<br>- **Links to other k-items:** References to the tensile testing machine, Fraunhofer IWM, and related experiments<br>- **Associated apps:** Applications for analyzing tensile test data<br>- **Semantic annotations:** Annotations providing context for the test conditions and results |
|---|---|
| **Data container** | CSV and HDF5 files containing raw and processed data from the tensile test |
| **Semantic graph** | RDF representation describing the tensile test data in a structured format while using standard terms (ontological concepts) for materials testing and mechanical properties. The RDF contains links to specific measurements and time series data stored in the data container. |

The concept of k-items in DSMS supports the philosophies of integration levels of semantic data and the DIKW (Data-Information-Knowledge-Wisdom) hierarchy that were introduced earlier. Each integration level represents a step towards a more comprehensive and meaningful understanding of the data, aligning with the progression from data to wisdom in the DIKW hierarchy. Table 3 illustrated how these concepts are related.

**Table 3.** Data integration levels supported by knowledge items and corresponding DIKW hierarchy levels

| Integration level | DIKW level | K-item |
|---|---|---|
| 0 | Data | The k-item contains basic metadata and raw data files. |
| 1 | Data | The k-item includes metadata describing the file type, enhancing basic data organization. |

| 2 | Information | The k-item includes metadata that provides contextual information and a semantic graph that contains basic relationships. |
| 3 | Information | The k-item metadata includes detailed descriptors (e.g., test conditions, machine details), and the semantic graph links this information. |
| 4 | Knowledge | The k-item's semantic graph links metadata with the data container, enabling integration and comprehensive data retrieval. |
| 5 | Wisdom | The k-item's semantic graph provides complete semantic annotations, facilitating advanced data integration, analysis, and decision-making. |

### 2.3.2. Knowledge item linkage

The effectiveness of our knowledge management concept, centered on notions of k-item and k-type, is significantly enhanced by the ability to link knowledge items (k-items) together. This linkage creates a rich, interconnected network of information, enabling users to navigate and explore complex relationships between different data points, datasets, and metadata. Knowledge item linkage is fundamental for establishing context, supporting advanced queries, and facilitating comprehensive data analysis. Linkage is achieved by connecting k-items either through simple relations, offering a basic level of integration, or through specifically defined relations within an ontology. More specifically, a default relation can be used to enable quick setup, generically stating that k-item A "is related to" k-item B. Alternatively, more expressive relations may be applied to model knowledge with higher precision, such as k-item A "is input for" k-item B.

Taking the example of the dataset shown in Table 2, in the following, we illustrate how k-item linkage works. The presented dataset includes tensile test data for sample XYZ, and its k-item is linked to several other k-items within DSMS, forming a network of related knowledge. The k-item for Dataset-001 includes links to the k-item representing the tensile testing machine. This linkage provides context by identifying the specific equipment used to generate the data, which is crucial for understanding the conditions under which the data was produced. For instance, knowing the machine settings, calibration details, and operational history can help in interpreting the test results accurately as well as validating them. The dataset is also linked to the k-item representing Fraunhofer IWM, the organization, at which the test was conducted. This connection places the data within the broader context of the institution's research activities, projects, and resources. It enables users to trace the dataset back to its source, facilitating collaboration and data sharing among researchers within and outside the organization. Additionally, Dataset-001 can be linked to other related k-items, such as datasets from similar experiments or complementary studies. For example, if there are other tensile tests conducted on different samples or under different conditions, these can be linked to provide a comparative analysis. This interlinking of k-items allows for a more holistic view of the material properties and performance. Furthermore, associated applications relevant to the k-item, such as software tools for analyzing tensile test data, are linked directly with the k-item. This integration allows users to seamlessly process and analyze the data using specialized tools, facilitating a streamlined workflow from data collection to knowledge generation.

Beyond simple linkages, more complex relationships can be expressed by k-item linkages, such as hierarchical structures. For instance, a k-item representing a research project can be linked to multiple datasets, each representing different experiments or phases of the project. These datasets can further link to individual data points, equipment, and personnel involved, creating a comprehensive hierarchy that mirrors the real-world structure of the research process. This hierarchical linkage allows users to navigate through different levels of data granularity, from broad project overviews to detailed experimental data, providing a clear and organized representation of the knowledge landscape.

### 2.3.3. Data integration: from raw data to information and knowledge

To integrate data into a dataspace, DSMS offers interfaces that enable flexible data integration through different types of modules, without being restricted to a specific technology for transforming raw data into semantic representations. To demonstrate this capability, we provide two modules: one for tabular data integration via a web form using the tool form2rdf, and another for integrating data files such as CSV and Excel using the tool data2rdf.

**Form2rdf: Semantification of web forms**

For enabling non-programmers to efficiently register information about different resources that constitute the knowledge graph, a user-friendly graphical user interface (GUI) is essential. Since DSMS provides a web-based GUI application, the use of web forms is a natural mechanism to implement this process, allowing users to input structured data through intuitive interfaces, such as web forms. The output of web forms is by design given as simple JSON objects. While JSON objects are convenient for manipulation, they are too simplistic to represent complex relationships between entities in a maintainable and scalable manner. To address this, additional processing is required to transform this information into robust and maintainable knowledge representations. This is where the tool form2rdf comes into play.

Form2rdf is specifically designed to enhance the capabilities of web forms within DSMS. This tool processes the data captured through web forms and transforms it into RDF format, thereby establishing connections between the data and other entities within the system's knowledge graph. The process, illustrated in Figure 4, starts with users filling out web forms embedded into k-items that were pre-defined by admins for the corresponding k-type. Once the data is submitted, form2rdf takes over, converting the JSON output into RDF. This conversion process involves mapping the form fields to ontological concepts, thereby creating subgraphs that represent the structured data. The data is automatically stored in the triplestore of the system.

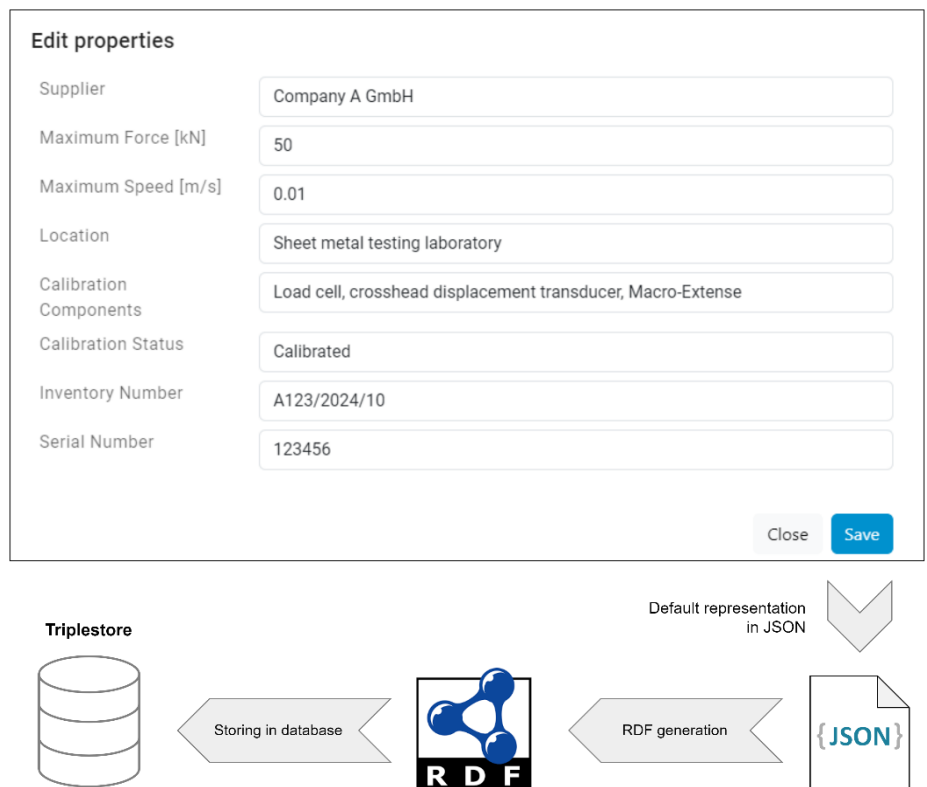

**Figure 4.** Data processing using form2rdf demonstrated with a testing machine. Users input relevant data, which is then transformed into an RDF representation and stored as a subgraph in the triplestore.

As well as filling out these web forms by users, defining them by admin users is facilitated through an intuitive interface, allowing this to take place without needing programming skills. Admins are provided with an additional console in the graphical interface to define the web forms for each k-type. In this console, admins can map the web form fields they define with concepts from ontologies curated in the system. This mapping information is used for the RDF generation when users submit the web form after inputting their data.

The generated subgraphs are the building blocks of the broader knowledge graph within DSMS. By linking these subgraphs with existing entities, form2rdf enables the formation of a comprehensive and interconnected knowledge graph. This interconnected structure serves as the foundation for representing and organizing heterogeneous data within DSMS, allowing for sophisticated querying, inference, and analysis across interconnected resources.

**Data2rdf: Semantification of raw data**

The integration of raw data into a semantic dataspace is crucial for enhancing data provenance, ensuring the reproducibility of research outcomes, and maintaining the reliability of experimental measurements and simulation results. This process is particularly important in fields where data files are generated by testing machines or simulations, as these files contain essential information about the processes and conditions under which the data was collected. For this purpose, we developed the data2rdf tool [43], an open-source Python library that facilitates the integration of data files into a dataspace. This tool allows users to upload raw data files, which are transformed to RDF (Resource Description Framework) and HDF5 representations. The RDF format is used to capture metadata, including details about the testing machine, the operator, the specimen's composition, and geometry, testing conditions, etc. These pieces of information are linked within the dataspace, enabling sophisticated querying of the knowledge graph, such as finding all datasets produced by a particular testing machine or listing all specimens made out of a specific material.

The dataflow of data2rdf, illustrated in Figure 5, requires a data file and a mapping file that links terms in the data file to ontological concepts. Data2rdf employs a modular design, supporting various data file formats, including Excel, CSV, and JSON. For Excel sheets, additional information is required that specifies the data locations within the sheet, and for JSON files, a JSONPath query is required. To accommodate more complex data structures, users can provide an additional RDF file to specify deeper RDF structures. If no such file is provided, a flat RDF structure will be generated. For storing tabular data, such as time series from experiments, HDF5 is used. Data2rdf generates RDF in a way that references HDF5 arrays, which represent columns of the original data table appearing in the input file, allowing users to download and further process the HDF5 data as a separate step.

Although data2rdf is a standalone tool, it can be associated with a k-item, triggering its functions whenever a file that meets specific criteria (as a simple example, by the file's extension) is uploaded. Both the raw data and the generated RDF are stored in the k-item for traceability purposes, with the RDF also being pushed into the system's triplestore. Designed for materials scientists with basic Python knowledge, data2rdf does not require extensive ontology expertise. Its user-friendly approach ensures that even those with minimal programming experience can effectively semantify their data. Through these capabilities, data2rdf ensures that raw data is transformed into structured, interconnected knowledge, enhancing the reusability and accessibility of scientific data within the dataspace.

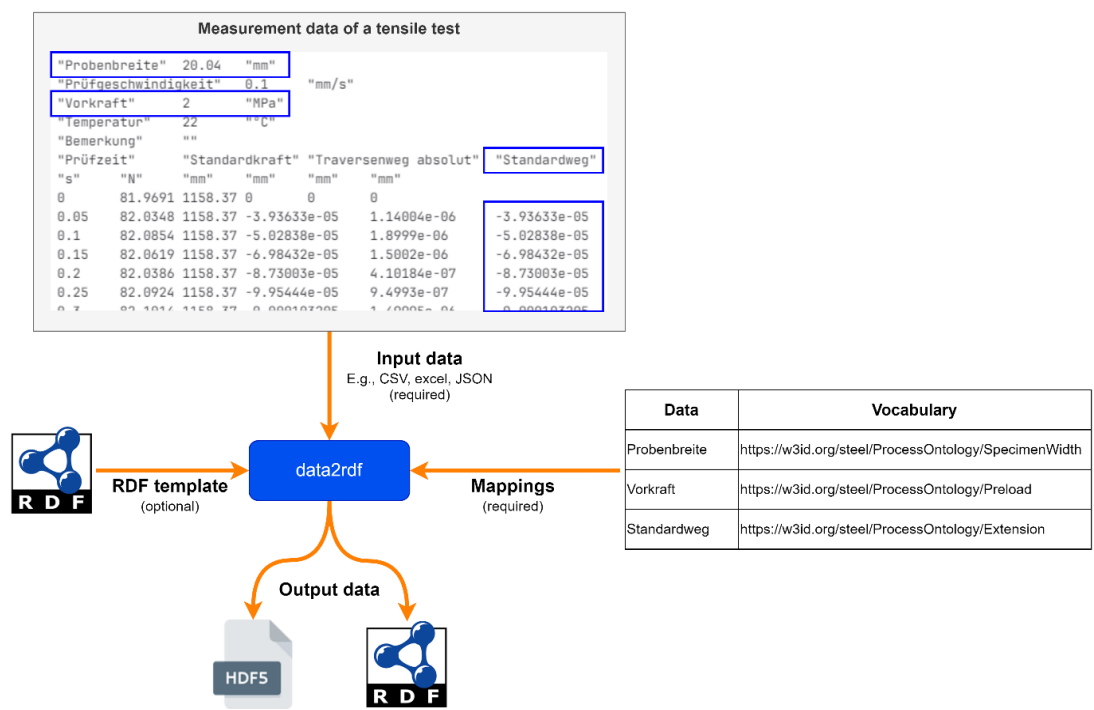

**Figure 5.** The dataflow of the data2rdf Python tool for a typical tensile test CSV file containing both metadata and time series data

### 2.3.4. Data exploration

The key capability of data exploration provided by DSMS enables users to efficiently locate, analyze, and understand the rich datasets and interconnected knowledge within the dataspace. The system offers multiple powerful tools and methods to facilitate this exploration, catering to both non-programmers and advanced, programming-affine users.

One of the main features for exploring data in DSMS is the free-text search for k-items. This search is enhanced by a language model that encodes the information of a k-item into a vector, enabling similarity searches [44]. Users can input general search terms and retrieve relevant k-items based on the semantic content of their descriptions, metadata, and annotations. This intelligent search mechanism makes it easier to find specific datasets, equipment, or research findings without needing precise keywords.

In addition to free-text search, DSMS offers faceted search for k-items. This allows users to specify search criteria such as the k-type and particular annotations, like a specific material type. Faceted search provides a more structured way to narrow down search results and find the exact resources needed. For instance, a user looking for tensile test data on a specific alloy can filter results by material type and k-item type, streamlining the discovery process. Once a k-item is found, users can explore its relationships with other k-items to uncover connections to various resources. For example, discovering the relationship between a dataset and the measurement machine that produced it can provide valuable context and insights.

This relational exploration is facilitated through visualization tools that display the networks resulting from linking k-items and from their semantic graphs, helping users understand the broader context and interconnections within the dataspace. For more advanced querying, DSMS includes a SPARQL console that allows users to perform detailed queries on the knowledge graph. While this tool may not be useful to all users, as it requires the user to have a deep understanding of SPARQL and to be familiar with the underlying ontologies, it enables more complex queries, leveraging the knowledge embedded in ontologies such as EMMO [29], BFO [30], and PMD core ontology [31].

### 2.3.5. Generation of new knowledge and recommendations

Going beyond data exploration, DSMS provides a platform for generating new knowledge through the development of specialized applications for data analysis and data processing. For this purpose, the DSMS Python SDK plays an important role offering a flexible and powerful interface for accessing and manipulating data stored within the dataspace. The Python SDK [45] allows users to explore the dataspace through scripts, making it easy to access data containers and semantic graphs of k-items. By leveraging popular Python libraries such as NumPy [46] and Pandas [47] to perform rich analyses and complex data manipulations or Scikit-Learn [48], TensorFlow [49] and PyTorch [50], to develop machine learning and data science applications. However, the user is not limited to using common Python libraries, but can, on the contrary, also apply complex (and potentially proprietary) software solutions, such as FEM simulation code for generating new knowledge [37].

Moreover, as it will be elaborated in Section 2.5, DSMS supports the integration of apps, developed by users, and additionally provides an app-store component in the system. Users with programming skills can write applications to perform specific tasks and publish these apps for others to use. This extensibility allows for the customization and enhancement of DSMS functionalities, making advanced analytical tools available to users with no or low programming skills. This collaborative approach ensures that the capabilities of the dataspace continually evolve and expand, driven by the contributions of its user community. By leveraging the semantic annotations and interconnected k-items within the dataspace, applications can automatically retrieve relevant datasets, apply predictive models, and present the results in an accessible format.

Introducing a fundamental development in DSMS, we present the novel concept of *semantic material cards* as an important example of knowledge generation. Typically, a material card is an input file that contains all relevant material properties and parameters for a specific material model used in numerical simulations. In contrast, a semantic material card is an abstract semantic representation of the material card that is stored in the dataspace as a subgraph. This subgraph contains the mechanical properties and material model parameters, not being limited to a specific material model. It is automatically generated through a parameter identification app that processes experimental measurement data. Therefore, semantic material cards are directly linked to the experiment(s) from which the mechanical properties and material model parameters originate. To obtain a syntactic (classic) material card, a second app exports semantic material cards into material input files, potentially for any desired simulation code on demand.

### 2.4. Ontology management

Ontologies are important components for creating an effective semantic representation of data to significantly enhance the FAIRness (Findable, Accessible, Interoperable, Reusable) of data. Proper management of ontology ensures that data is annotated with meaningful, standardized terms, facilitating better findability and interoperability across datasets and applications.

Semantic annotations, as shown in Figure 3, are used to label k-items. These annotations go beyond the simple hashtag systems commonly found in social media by providing a more structured and manageable way of labelling information. By using controlled vocabularies, k-items can be systematically tagged, making it easier to locate relevant datasets and resources. Also, raw data, such as measurement data, is enriched with the system's ontology. This enrichment allows for an easy identification and retrieval of specific data points, such as columns of tabular data. This ensures that even the granular aspects of data are searchable and integrable within the broader knowledge framework.

The ontology in DSMS is both customizable and expansive, incorporating specific domain terms while also leveraging widely established vocabularies such as DCAT [51], DC-terms [52], and QUDT [53]. Domain experts are provided with means to enter new terms into the system's vocabulary, which are then made available to all users. This dynamic process may be facilitated through a pipeline for registering new terms via a REST API, allowing various tools to plug into the system and offer this functionality. One instance of such a pipeline integrated into DSMS has been implemented during the project StahlDigital [37]. The advantage of the adopted procedure is that a domain expert can easily add a vocabulary or glossary term. Therefore, the barrier for integrating data with new terms is very low.

Typically, when we refer to vocabulary, we mean a controlled list of terms with either no hierarchy (a flat list) or a simple hierarchy (taxonomy). However, DSMS also supports more complex ontologies by connecting the vocabulary terms to each other with meaningful relations. This approach aligns with the integration levels described in Table 1, enabling an incremental approach for the semantic level of the system. Also, the vocabulary is instrumental in determining the compatibility of resources with each other within the system. For instance, a script designed to render time series data can be automatically made available for a k-item of type dataset if the k-item annotation indicates that it contains such data. This principle will be extended in the future to enable semantic workflows, where compatibility between apps can be determined and integrated seamlessly, creating a coherent and efficient data processing pipeline.

### 2.5. App store and workflows

An app within the DSMS ecosystem is a software component created by users and made available to others through an app store or virtual marketplace. This concept is particularly relevant for the materials modelling community, where simulation engines such as LAMMPS play a central role. Ideas about such marketplaces have been explored in EU initiatives like Materials MarketPlace [54, 55], VIMMP [56], and DOME 4.0 [57], underscoring their importance. Simulations in materials science are crucial for several reasons:

- **Innovation and regulations.** They help save resources by meeting complex sustainability requirements for materials, processes, and products, and address societal needs for ecological and sustainable solutions.

- **Fragmented expertise.** The manufacturing industry often requires specific expertise, which is scattered across a fragmented market of providers.
- **Accessibility barriers.** Newcomers face hurdles in accessing software tools, with specialized expertise and support often being prerequisites.

These factors highlight the need for a platform solution that creates new opportunities for innovation in sustainable materials, manufacturing, and products. A virtual marketplace reduces barriers to the use and exchange of data, applications, and knowledge related to materials.

In DSMS, applications can be hosted on a Kubernetes cluster, where they are provided as Docker images. Simple use cases for these applications include processing an input file to generate an output file that can be rendered on the frontend and downloaded for further processing. Another example is a visualization app that renders input data. Additionally, an app may serve as a gateway to more complex simulation engines, providing a tailored user interface for specific use cases where the user can adjust control parameters.. DSMS also offers tools for non-programmers to define web forms for these applications, eliminating the need to develop a separate frontend. Apps can also be configured to trigger automatically, which is particularly useful for tasks such as data validation, data integration, and post-processing.

Apps in DSMS can function as standalone tools or communicate with each other to form workflows, which are especially beneficial in materials science and production, where complex, multi-step processes are common. To support this, the DSMS technology stack employs Argo Workflows [58], an open-source, robust and popular workflow engine. Argo enables efficient orchestration of these workflows, automating tasks such as data preparation, analysis, and reporting. Semantic technology plays a crucial role in these workflows, ensuring data interoperability between apps. For example, the output data from one app can potentially be used as input for another without requiring a developer to manually integrate the two apps (this capability has yet to be fully implemented though). Finally, admin and user consoles provide control over app visibility and access, with system monitoring in place to ensure secure communication.

## 2.6. Integrating external data repositories

Valuable data is often stored in external data repositories such as Nomad [59], Zenodo [60], Figshare [61], and CKAN [62], which provide a wide range of functionalities for managing and sharing datasets. Initial versions of dataspace connectors for such repositories have been integrated into DSMS with built-in mechanisms. While still in a relatively early stage, this capability represents a core function of DSMS, with plans to expand it further and incorporate external connector technologies like EDC [63, 64].

While the aforementioned systems are effective in cataloguing and making datasets accessible, they have some limitations:

- **Limited semantic capabilities.** These repositories focus primarily on cataloguing and making datasets accessible, but they lack advanced semantic capabilities.
- **Restricted data integration.** They are designed for publishing and sharing datasets but offer limited tools for integrating diverse data types and formats.
- **Basic user interaction and analysis tools.** They typically provide basic features for dataset visualization and metadata management but lack sophisticated tools for data exploration and analysis.

DSMS complements the functionalities of typical data repositories through its core competencies, as discussed earlier. While the data remains in the original repositories, DSMS creates metadata and semantic representations, enhancing the usability and integration of the data. The migration process from external repositories to DSMS is currently performed periodically via an automated workflow. This ensures that the metadata and semantic representations in DSMS are up to date with the original data sources. In the future, admin consoles will be provided to offer better control over the connected repositories, allowing for more efficient management and integration of external data.

## 2.7. Architecture of DSMS-powered dataspaces

The architecture of DSMS-powered dataspaces, as illustrated on a high-level in Figure 6, employs a microservice-based approach to provide flexibility, scalability, and efficient data management. Each component within the system serves a specific role, ensuring modularity and smooth operation. This subsection also serves as a summary of the key architectural elements described throughout Section 2, highlighting how the individual services and components work together to form a cohesive and integrated system. Key components of the architecture include:

- **Frontend (Angular).** A responsive, user-friendly interface built with Angular, allowing users to interact with the system via RESTful HTTP calls to the backend services.
- **Public API (NGINX reverse proxy).** A secure entry point for accessing the system's services, facilitating communication between the frontend, user-defined apps, and backend services.

- **Knowledge services.** The knowledge service manages the k-items within the system and the knowledge type service handles k-types, providing a dynamic mechanism to classify and customize the knowledge items.
- **Vocabulary service.** Supports the management of the system's vocabulary, including registering new terms, searching existing ones, and connecting terms to ontologies (e.g., the PMD core ontology).
- **User service.** Provides user management functionalities, built on top of Keycloak [65, 66] for secure authentication and user management.
- **Integrated Jupyter environment.** Enables advanced data analysis and exploration through Python-based workflows. Users can share scripts and collaborate within the system.
- **Data storage.** PostgreSQL databases are used for storing structured data related to knowledge items and user management, while a triplestore stores semantic data, enabling sophisticated querying and linking to ontologies.
- **Workflow engine.** Utilizes Kubernetes and Argo to manage data integration pipelines, ensuring efficient execution of complex tasks and workflows.

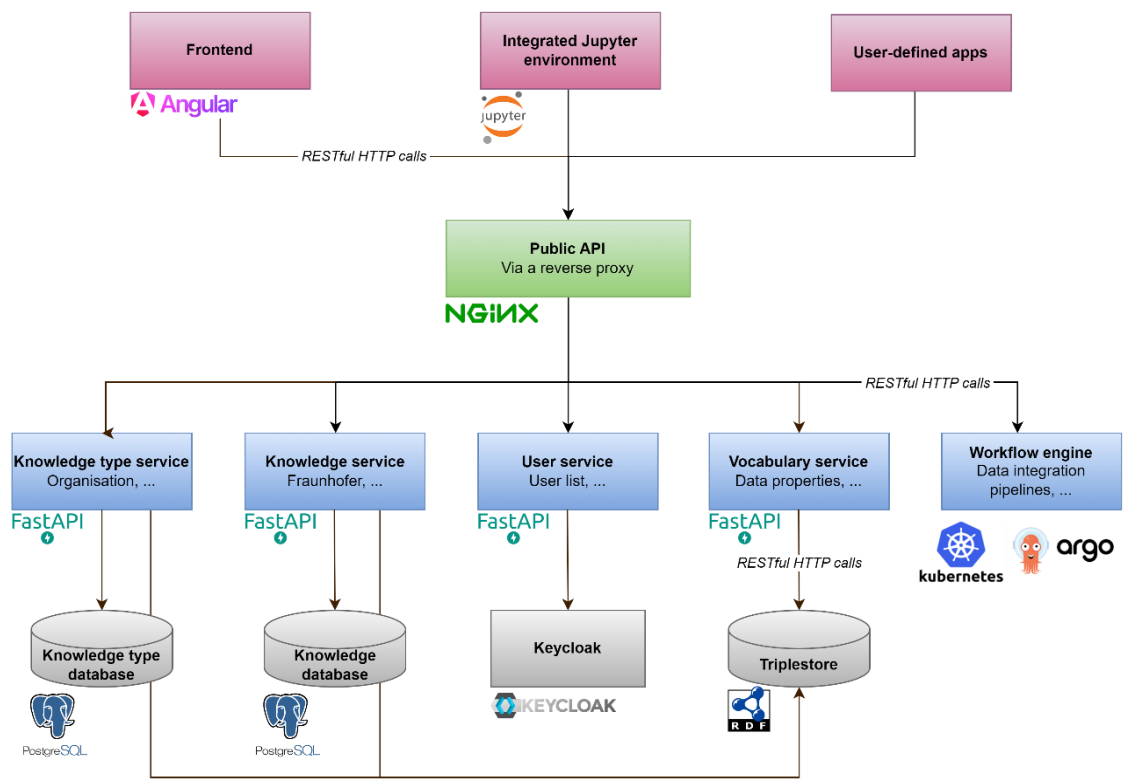

**Figure 6.** High-level system architecture overview: integration of frontend clients, API, and backend services with workflow engine and data and vocabulary (ontology) management components

All backend services are developed using the FastAPI framework, ensuring high performance, scalability, and ease of integration. Figure 6 presents this high-level architecture, illustrating how the components interact through RESTful HTTP calls and are integrated to manage knowledge, user data, and workflows within the DSMS ecosystem.

## 3. Demonstration and validation of DSMS-based dataspace solutions

To validate the applicability and demonstrate the benefits of the knowledge management concepts presented in this paper, we consider various materials, testing methods, and data sources. The application cases involve steel and copper materials from the research projects StahlDigital and KupferDigital, both part of the German MaterialDigital initiative. These two use cases represent typical scenarios in which we illustrate how addressing data management aspects through a DSMS-powered dataspace solution is both simple and effective. The key benefits of using DSMS include: facilitated handling and integration of heterogeneous data, representation of data processing workflows in a FAIR manner, data exploration and analysis, and the derivation of recommendations and decision support, all within a single comprehensive technology stack. The capabilities of DSMS, as described in Section 2, extend to managing data throughout the entire value chain of a material and the lifecycle of the related product. This includes the mining of raw materials, material synthesis, production of semi-finished products and components, component usage, and end-of-life/ recycling.

### 3.1. StahlDigital: Demonstrating data integration, data exploration, and knowledge generation

In the context of research project StahlDigital, we demonstrate and validate key functionalities of the dataspace powered by DSMS. These functionalities include the integration and linkage of different types of data, data exploration, and the generation of new knowledge. It is important to note that DSMS enables data integration at all the interoperability levels described in Table 1, providing a comprehensive and flexible solution for managing diverse datasets.

#### 3.1.1. Data integration

The core idea of our data management concept is to transform relevant metadata from a data source to a semantic level, while making the data itself, such as time series, easily accessible to both computers and humans. This approach enables sustainable and efficient use of the integrated data (e.g., evaluating and visualizing data promptly using the ontology-based harmonized data representation). Without such integration, these tasks would be repetitive, ad-hoc processes, which are time-consuming and error-prone.

During the research project StahlDigital, data from various experimental sheet metal testing methods (including tensile tests, notched tensile tests, shear tests, bulge tests, Nakajima tests, and component crash tests) was gathered and integrated into the StahlDigital dataspace solution. The raw measurement data was available in the form of CSV or Excel files. As a basis for data integration, relevant ontological concepts for these experiments were collected by domain experts and made permanently and publicly available [37, 67]. Using a tensile test with measurement data in CSV format as an example (see Figure 7, top left), we demonstrate how data is integrated using the data2rdf tool. Data2rdf essentially requires two input files:

1. Mandatory: A mapping file, provided by a domain expert, that relates terms used in the measurement file to the defined ontological concepts. This can be a simple two column file (column one for the terms, and another for the corresponding ontological concepts) or a structured JSON file, as depicted in Figure 7 (top right). In this case, the JSON mapping file relates metadata from the header of the tensile test measurement file to corresponding ontological concepts.

2. Optional: An RDF graph can be provided to specify deeper relationships between the ontological concepts defined in the mapping file. If no RDF graph is provided, data2rdf generates a flat RDF graph, as depicted in the top part of Figure 8. The bottom part of the figure illustrates how generated graphs can be enriched by incorporating additional knowledge using this parameter, as demonstrated by leveraging the PROV ontology [68] in this example.

**Figure 7.** Header of a tensile test file integrated into the StahlDigital dataspace using data2rdf tool (left). The necessary JSON mapping file (right) relates the terms from the header to corresponding ontological concepts.

Units stated in the experimental measurement files are mapped using the QUDT ontology [53], allowing error-free unit conversion. The mapping file allows to map a value to a vocabulary term when executing the data2rdf tool. In this example, the material specified in the tensile test header is concatenated with the annotation prefix specified in the mapping file to create the corresponding term in the ontology, as shown in the third entry in Figure 7 (top right). This will result in the ontology term *https://w3id.org/steel/ProcessOntology/DX56D* in this case.

Through this integration procedure, more than hundred datasets coming from the aforementioned testing methods were integrated into the StahlDigital dataspace. Since new ontological concepts can be added via the integrated vocabulary service, the barrier to integrating new types of data is significantly low. In StahlDigital, the data was integrated according to Table 1 at interoperability level 4 (meta data was completely semantified, while the time series data was integrated as an HDF5 container for efficiency). Additionally, the original raw data file was uploaded to the same k-item for reproducibility purposes (interoperability level 0).

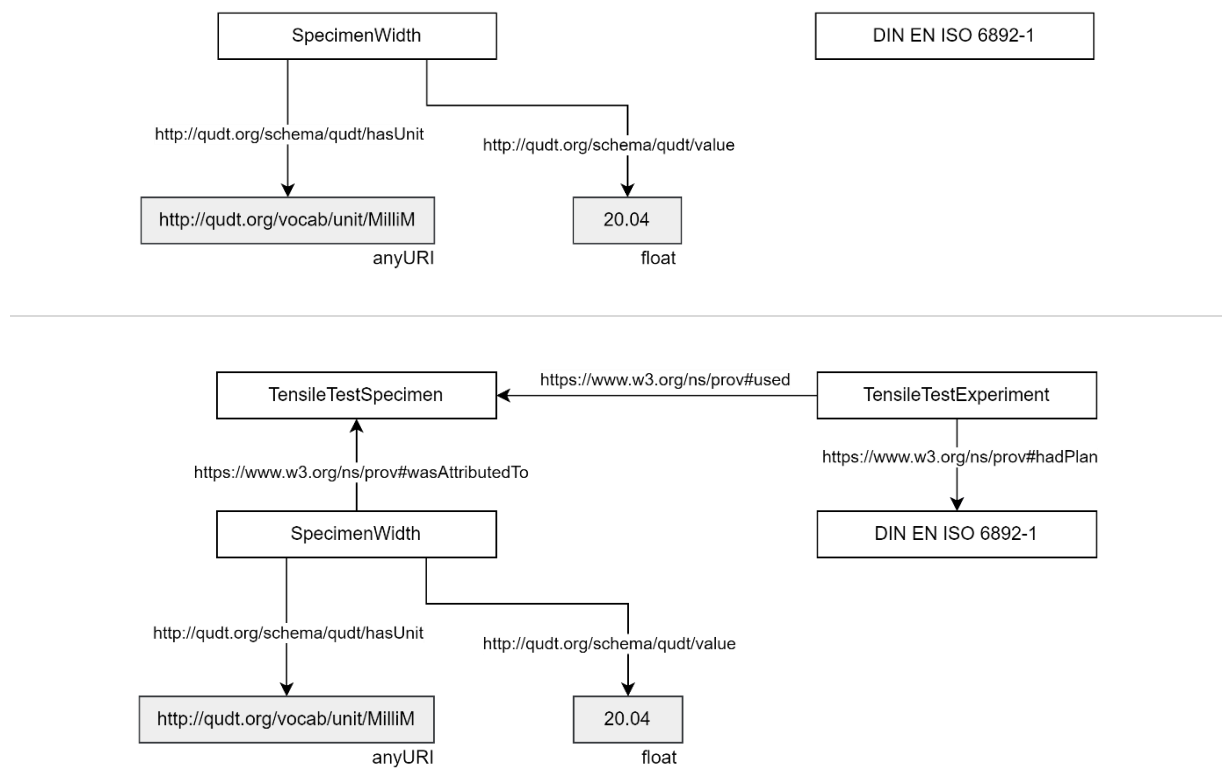

**Figure 8.** Generated subgraphs are shown without (top) and with (bottom) the optional RDF graph parameter. Including this parameter allows the incorporation of a pre-defined ontology template, enabling the creation of more complex, interconnected graphs. In contrast, its absence results in a default flat list of ontology terms.

### 3.1.2. Data exploration

Annotating k-items is particularly useful for searching the dataspace. For example, when searching for tensile tests conducted with a certain material, as illustrated in Figure 9 (top), users can quickly locate relevant k-items based on specific criteria. The outcome of searching for the ontological concepts *TensileTest* and the material *DX56D* in the frontend using the search functionality is shown in the figure.

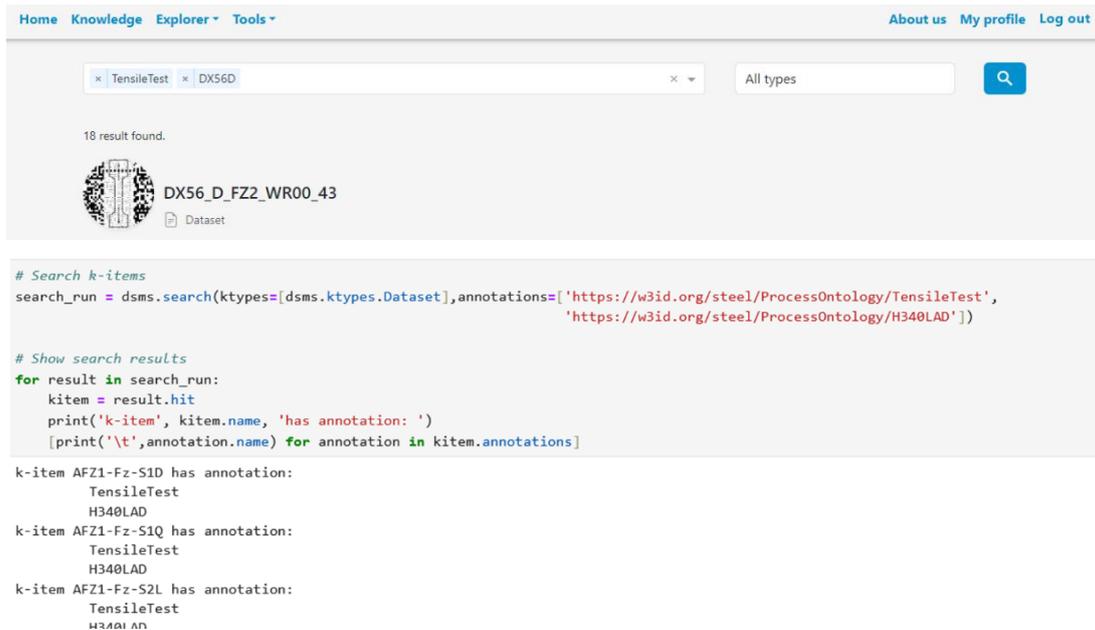

**Figure 9.** Frontend view for searching the dataspace for tensile tests with DX56D material (top). DSMS-SDK example for searching all tensile test data with material H340LAD (bottom).

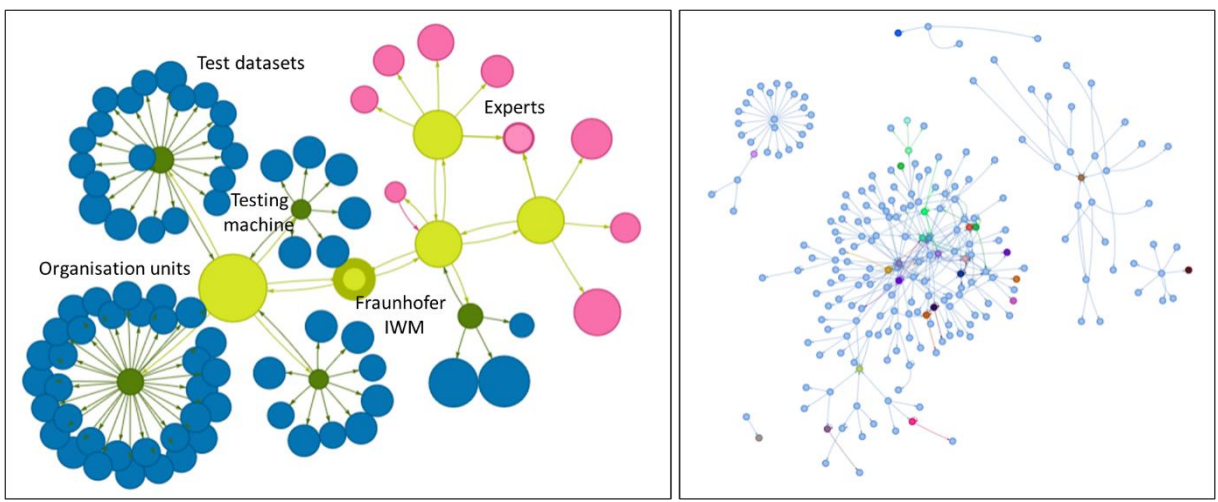

**Figure 10.** Graph representations of the k-items linked to the center node Fraunhofer IWM in the StahlDigital dataspace (left) and a complex RDF graph of an integrated tensile test (right).

The same search functionality is provided via the DSMS-SDK. This Python interface enables users to process data according to the specific needs of a use case. Figure 9 (bottom) provides an example of how the DSMS-SDK can be used to programmatically retrieve and process data, offering flexibility and customizability for advanced data analysis and application development. Through the SDK related data can also be retrieved in standard formats such as HDF5, pandas DataFrames, or RDF, which allows further processing using compatible Python libraries as described. The DSMS frontend further enhances data exploration with robust data visualization capabilities. Users can view key-value or key-value-unit data, time series data, annotations and links to other k-items (Figure 10, left), as well as the underlying RDF subgraph of the integrated data (Figure 10, right).

**Generation of new knowledge**

After data visualization and analysis, engineers typically post-process the data to gain new insights, such as understanding the behavior of a material. Such workflows can be easily set up and automated via DSMS, ensuring that the newly generated knowledge is stored semantically and can be traced back to its origins. In the following, we demonstrate such a workflow at the example of the semantic material card concept, described in Section 2. Specifically, we show how integrated primary (raw) data from mechanical tests can be further processed and prepared for use in numerical simulations (see Figure 11 for an illustration of this workflow).

When new tensile test data is fed into the StahlDigital dataspace, a tensile test evaluation app automatically initiates. Due to the semantic data integration, the original format of the measurement data does not affect this workflow. The tensile test evaluation app identifies material model parameters and mechanical properties, such as Young's modulus, yield strength, and parameters for the Hockett-Sherby model. In addition to parameter identification, the app generates a report and stores both the identified parameters and the report in the dataspace. To ensure reproducibility, the settings for the tensile test evaluation app are also stored. The identified parameters, stored in a semantic manner, form the semantic material card. This information is used by a second app to export syntactic material cards, for example, for finite element software such as Abaqus or LS-DYNA. The representation of this workflow in the dataspace forms a graph, in which different k-items are linked, as shown in Figure 12.

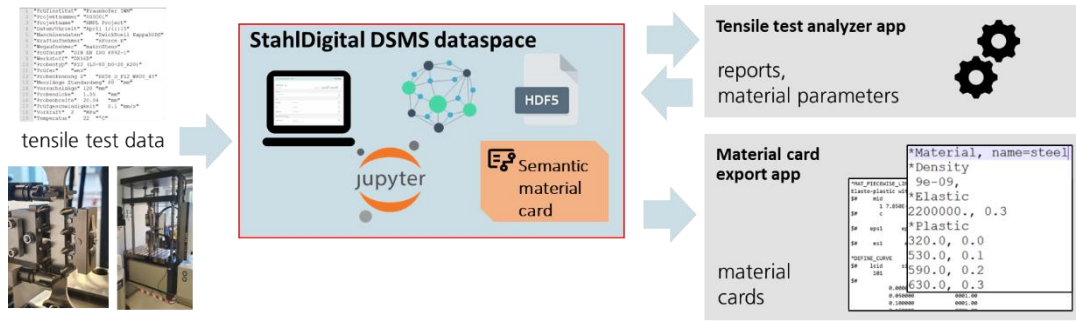

**Figure 11.** Data processing workflow of the semantic material card use case in the StahlDigital project

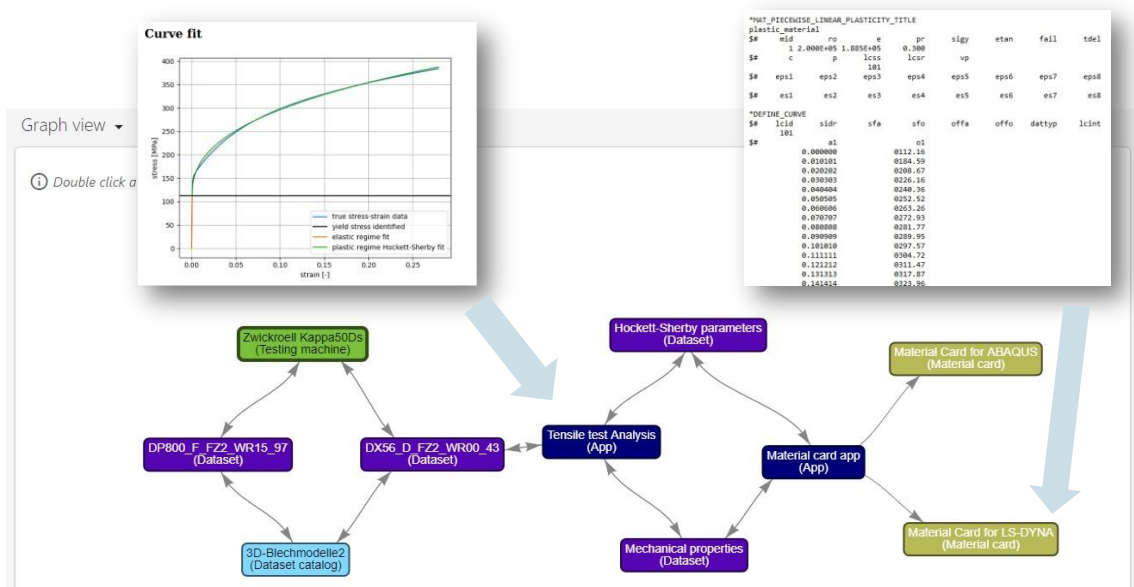

**Figure 12.** Graph resulting from the data processing workflow of the presented StahlDigital example

### 3.2. KupferDigital: Alloys design example using hardness measurement data

In the second use case, we demonstrate how DSMS can help in addressing challenges in materials design processes, such as the development of new alloys, in a generic and efficient manner. As illustrated in Figure 13, the primary advantage of using DSMS in the context of data-driven materials development is its ability to integrate all steps in digitalization and scientific work into a single interactive framework. This comprehensive approach encompasses data and workflow documentation, data aggregation and integration into a dataspace, querying the dataspace, and final data analysis. By unifying these processes within one framework, DSMS maximizes efficiency in terms of time and resources.

Specifically, the goal of the materials development use case in KupferDigital was to derive a copper alloy with high material strength for construction purposes. Since hardness corelates nonlinearly with material strength but is much simpler to measure, the subset of relevant copper alloys achieving higher strength is identified by the Brinell hardness value. In a pre-screening process step, tensile strength is measured for each investigated copper alloy. The subsequent step in the materials development strategy, detailed in separated reports [38, 69], involves aggregating information about a specifically chosen alloy using DSMS. This analysis provides insights into the correlation between alloy strength, or hardness, and the corresponding alloying elements, such as Ni, Al, Si, Sn, or Zn.

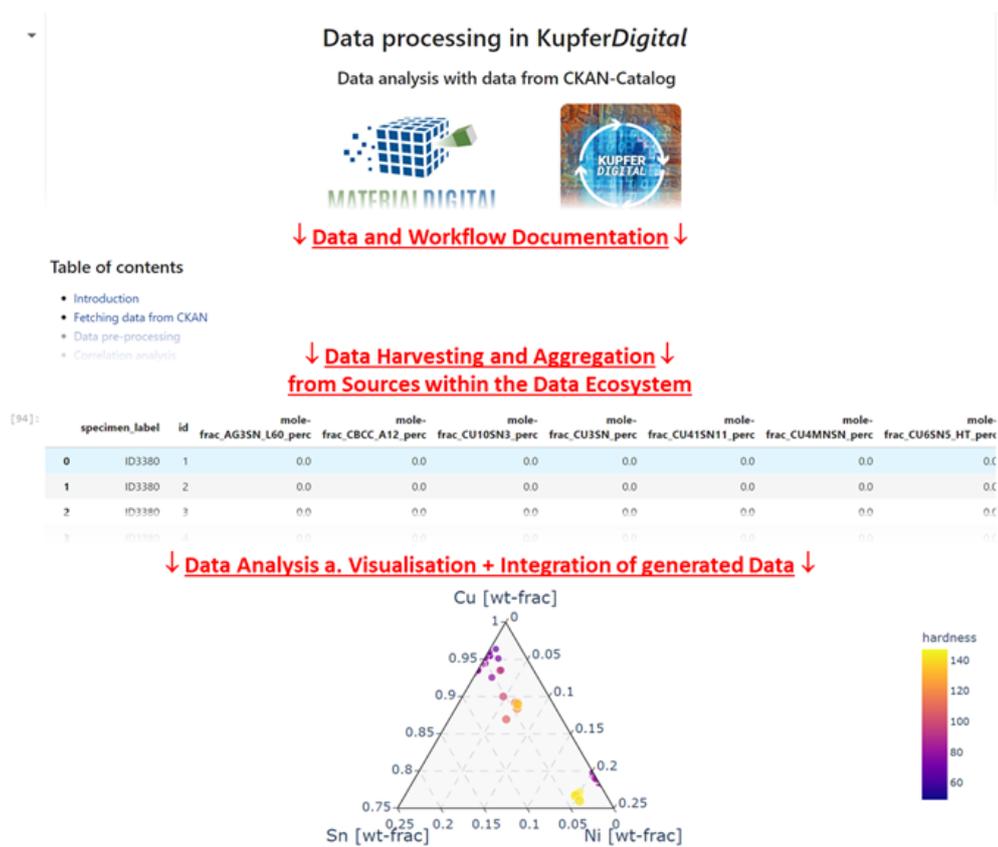

**Figure 13.** In KupferDigital, DSMS was used as an interactive platform for data-driven materials development. The analysis concept shown here was developed in project CuBiK [69] and was to be realized with semantic technologies in KupferDigital [38].

### 3.2.1. Data integration

In KupferDigital, hardness data was measured for a variety of copper alloys by the project partner Bundesanstalt für Materialforschung und -prüfung (BAM). A total of 30 datasets were integrated into KupferDigital dataspace, covering six different copper alloys: CuZn38As, CuZn21Si3P, CuNi12Al3, CuSn12, CuSn6, and CuNiSi, with five repetitions for each experiment. The integration of the data was performed using the data2rdf tool, as described in the StahlDigital example, with minor adaptations to accommodate the BAM data. One of the measurement files in CSV format for the CuZn38As copper alloy and the corresponding mapping file both are depicted in Figure 14.

Additionally, these datasets were semantified and integrated by BAM into a CKAN-based data repository (ckan.kupferdigital.org) and published via Zenodo [70]. DSMS facilitates the integration of data from the CKAN repository by periodically polling CKAN and creating corresponding k-items in the dataspace. It is important to note that the data remains on the external repository and only being referred to by DSMS. This approach avoids the need for heavy-weight data migration while still enabling the full benefits of DSMS.

```
 1  Name,Value,
 2  ID,1.0,
 3  Test Piece Identifier,A,
 4  Test Piece Composition,CuZn38As,
 5  Test Piece Producer,Copperalliance,
 6  Indentation Repetition,1.0,
 7  Indentation Horizontal Diameter,0.86316,mm
 8  Indentation Vertical Diameter,0.83682,mm
 9  Indentation Average Diameter,0.84999,mm
10  Brinell Hardness,106.89333758774,HBW
11  Total Average Diameter,0.833716,mm
12  Average Brinell Hardness,111.476225984258,HBW
13  Standard Deviation of Brinell Hardness,6.71000409814515,HBW
14  CRM Average Brinell Hardness,141.390453006151,HBW
15  CRM Standard Deviation Brinell Hardness,0.549079651735196,HBW
16  CRM Uncertainty (UCRM),0.995,UnitOne
17  Testing Machine Uncertainty (Uh),0.625950802978124,UnitOne
18  Measurement Resolution Uncertainty (Ums),0.123289022262944,HBW
19  Permissible Uncertainty (Umpe),4.26,UnitOne
20  Brinell Hardness Uncertainty,5.45755974433978,HBW
21  Test Piece Thickness,8,mm
22  Test Piece Processing,casting and rolling,
23  Test Piece Preparation,smoothing polishing and cleaning,
24  Indenter Identifier,3688.0,
```

```
 1  {
 2    "ID": {
 3      "key": "ID",
 4      "iri": "https://w3id.org/steel/ProcessOntology/Identifier",
 5      "annotation": ""
 6    },
 7    "Test Piece Identifier": {
 8      "key": "Test Piece Identifier",
 9      "iri": "https://w3id.org/steel/ProcessOntology/TestPieceIdentifier",
10      "annotation": ""
11    },
12    "Test Piece Composition": {
13      "key": "Test Piece Composition",
14      "iri": "https://w3id.org/steel/ProcessOntology/TestPieceComposition",
15      "annotation": ""
16    },
17    "Test Piece Producer": {
18      "key": "Test Piece Producer",
19      "iri": "https://w3id.org/steel/ProcessOntology/TestPieceProducer",
20      "annotation": ""
21    },
22    "Indentation Repetition": {
23      "key": "Indentation Repetition",
24      "iri": "https://w3id.org/steel/ProcessOntology/Repetition",
25      "annotation": ""
26    },
27    "Indentation Horizontal Diameter": {
28      "key": "Indentation Horizontal Diameter",
29      "iri": "https://w3id.org/steel/ProcessOntology/IdentationDiameterHorizontal",
30      "annotation": ""
31    },
32    "Indentation Vertical Diameter": {
```

**Figure 14.** Hardness measurement file (left) and a corresponding mapping file (right)

### 3.2.2. Data exploration and knowledge generation

After integrating the data into the KupferDigital dataspace, hardness measurements were analyzed interactively using the DSMS Python interface and the DSMS-SDK. This interactive approach allows for the immediate generation of new insights, even from initial data visualizations. For demonstration purposes, the measured Brinell hardness was plotted against the individual copper content of each alloy in the form of a violin plot, showing the distribution of the measurements (see Figure 15). The analysis reveals, that CuNi12Al3, with the alloying elements nickel and aluminum, achieves the highest Brinell hardness at an intermediate copper content of about 85 percentage by weight (wt.%). Higher copper contents, while expected to improve conductivity due to the electrical properties of copper, result in lower hardness for the alloys caused by the low mechanical strength of it, as indicated by the trend curve in Figure 15. Conversely, lower copper content leads to higher Brinell hardness for the studied alloys, though with different alloying elements.

This, however, comes at the cost of reduced conductivity, explained by the relationship between the copper content and the electrical conductivity of copper alloys, which is dominated by copper content in a composition. Interestingly, both CuZn38As and CuNi12Al3 alloys deviate from the identified trend, from which we can derive a recommendation for further analyses to understand the mechanisms behind these anomalies. Such deviations suggest that additional factors may influence the hardness and conductivity of these alloys, which must be considered in subsequent alloy development steps. In summary, by leveraging DSMS for data exploration and analysis, researchers can efficiently visualize and interpret complex datasets, facilitating the discovery of critical insights that drive material innovation. This capability not only enhances the understanding of material properties but also informs the strategic development of new, optimized alloys. While hardness is one important mechanical property in the development process of copper alloys, the further reaching potential of the KupferDigital dataspace will be evaluated in future for targeting other functional material properties or also modelling concepts.

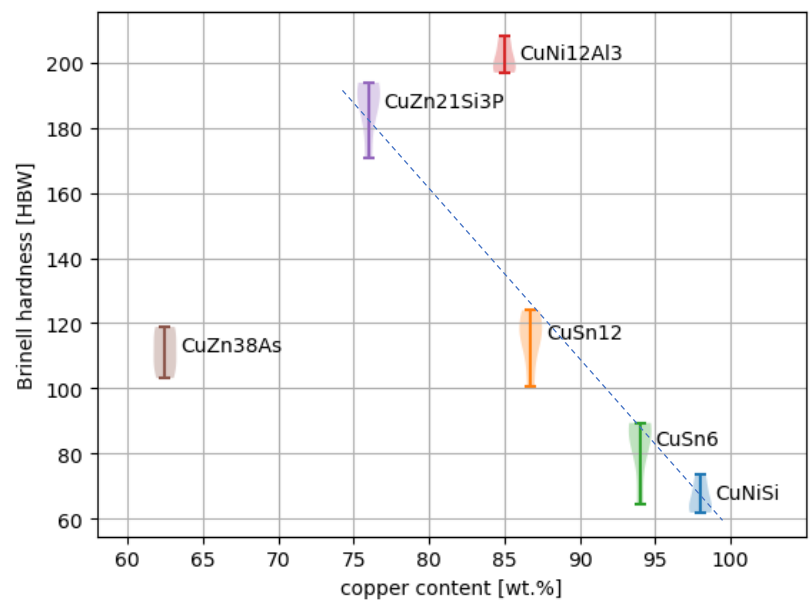

**Figure 15.** Brinell hardness vs. typical copper content of the investigated alloys in the research project KupferDigital. The dashed line represents a trend visible in the plot.

## 4. Discussion

The application cases described above provide an initial insight into the capabilities of DSMS-powered dataspaces, while demonstrating its basic functionalities, namely, data integration, data exploration, generation of new knowledge, and recommendations. The presented use cases highlight the potential of DSMS to transform raw experimental data into actionable insights and facilitate complex workflows in materials science and manufacturing. With respect to our research hypothesis, we have demonstrated that DSMS offers a viable solution for overcoming the challenges of modern materials and process data management, adhering to FAIR data principles through the use of semantic technologies.

In contrast to typical top-down approaches (see Section 0 and corresponding efforts [71, 38]), DSMS provides functionalities to integrate data in a semantic manner without the need of an all-explaining ontology. Nevertheless, one of the key strengths of DSMS is its ability to support complex data processing workflows. Only a mapping file (and, in case of Excel-files, an additional file specifying the location of the quantities) is required to transform any experimental raw data into RDF and integrate it into the dataspace in a FAIR manner. This capability not only streamlines the processing of raw data but also ensures that the resulting knowledge is stored in a traceable and semantically enriched manner. It also enables efficient semantification and storage of large amounts of heterogeneous data in an automated way.

The bottom-up approach of DSMS is also relevant for other functionalities beyond data integration. For example, in translation processes where industrial problems, such as those from production engineering in SMEs (small and medium-sized enterprises), are translated into the material science context. This translation can occur either automatically or through the efforts of domain experts acting as translators. One of the key outcomes of the European project OntoTrans [72] was to demonstrate how semantic technologies could assist in this process. However, a top-down approach in this context also presents a bottleneck, hindering problem owners and translators from effectively engaging with the core translation tasks. We argue that the DSMS philosophy, with its bottom-up approach, will also benefit these activities by facilitating a more flexible and user-friendly translation process.

When it comes to analysis and exploration of integrated data, DSMS enables core functionalities within one system (see Figure 1), which is a notable advantage over commonly used data repositories and management systems such as CKAN [62]. In addition, DSMS offers interfaces for users of different backgrounds. For example, it provides a SPARQL console for ontology exerts but also more user-friendly and intuitive tools, such as the free-text search for non-programmers, as shown in the StahlDigital use case. The provided visualization tools help engineers to understand the data, gain new insights and derive recommendations for further development steps without the need to switch systems. This is illustrated in the KupferDigital use case at the example of copper alloy design. With only few lines of code, data found using the search functionality can be plotted and analyzed easily (independently of the amount of data) using the DSMS-SDK. This process, although straightforward, is time-consuming without the use of semantic technologies, as it requires manual data collection, transformation into a usable format, plotting and potentially switching between tools. It is worth noting that without the techniques provided by DSMS, this process becomes increasingly labor-intensive as the amount of data grows, making it difficult for engineers to maintain an overview.

Furthermore, due to the enabled interoperability by the semantic representation, more powerful tools like machine learning models or simulation software can be seamlessly coupled with the dataspace to automatically investigate new relations between materials data (e.g., process-structure-properties-performance relations) or to enrich the data with new knowledge. Such a more complex workflow example realized on top of DSMS as presented in [37]. Graphs resulting from such workflows underly typically a large complexity, as can be seen in Figure 12. Although the workflow in the StahlDigital use case does not incorporate a complex material model and a vast amount of different experimental measurements typically needed (e.g., for crystal plasticity simulations), the resulting graph is already quite intricate. Logging all this information to keep track of the information flow without semantic technologies is challenging if not even impossible for a human being alone (this includes e.g., to note which models and data was used, which software settings were defined, who conducted the experiment und which conditions, which decisions were made and why, etc.). In contrast, with the help of DSMS, data provenance can be maintained on the fly, hidden from the user, thereby reducing the overhead. Reproducibility of results can be guaranteed easily, ensuring that the entire data processing pipeline is transparent and traceable.

In contrast to other dataspace initiatives like Gaia-X [73], Catena-X [74], and the International Data Space (IDS) [36], DSMS is designed to instantiate multiple dataspaces tailored to well-defined user groups. This flexibility enables the setup of internal dataspaces for a single organization or department, collaborative dataspaces across multiple companies, and fully public dataspaces (e.g., as outlined in [26]). While other initiatives also aim to facilitate secure and standardized data exchange across organizations, DSMS focuses primarily on advanced data and knowledge management functionalities. Future versions of DSMS will potentially incorporate connector technologies from broader dataspace initiatives such as IDS, Gaia-X, and Catena-X to enable integration with larger data ecosystems, while also supporting essential data sovereignty features.

A special feature of DSMS in contrast to other dataspaces is given by the fact that the data in the dataspace can be enriched via applications running or initiated on the DSMS technology itself. To our knowledge, such a feature is rare in other dataspace concepts. This feature enables enormous possibilities for data processing applications, such as artificial intelligence applications and other types in the framework of ICME [1].

Nevertheless, at the current development level, DSMS has some limitations that need to be addressed in the future. One of these is the requirement for periodic synchronization with external data sources, meaning that changes in an external database are not immediately reflected in a DSMS-powered dataspace. Additionally, the data exploration functionality in external sources is limited to simple text search. Furthermore, although the vocabulary service is comprehensive, ontology curation still needs to be carefully performed manually by an ontology expert, which can become costly for a general usage of DSMS. AI can potentially assist with this, but the application of AI (specifically large language models) to enhance DSMS has not yet been fully investigated. AI could also aid in achieving higher levels of integration, which can be a labor-intensive challenge (e.g., making an app seamlessly communicating with another app).

Moreover, the deployment of DSMS-agnostic secrets for external data sources is not fully developed, leading to potential security risks if users manually input authentication secrets. Best practices recommend using a central Lightweight Directory Access Protocol (LDAP) to integrate the identity management of DSMS and external data sources, thus preventing leaks of confidential data. This approach would generally improve security and prevent any data breaches within the system.

## 5. Conclusion

In this article, we presented novel concepts and their evaluation for (i) material data integration and linkage, achieved by harmonizing data representations through semantic technologies, (ii) effective exploration of the semantically enhanced material data, (iii) knowledge generation and management on a semantic level, and (iv) knowledge graph-driven recommendations. These functionalities are at the core of DSMS, a methodological framework realized as a technology stack, offering a powerful approach to data and knowledge management rooted in semantic technologies and designed to enable FAIR data principles. This methodology represents a significant advantage over traditional data management solutions, which typically stop at the information level. Furthermore, while classic data management solutions generally require users to conform data to a predefined format, DSMS provides the flexibility to integrate virtually any type of data. The semantics-driven concepts for orchestrating data and transforming it into information and knowledge, along with derived recommendation capabilities, have the potential to revolutionize the engineering processes for designing materials, manufacturing processes, and products, while managing heterogenous data in complex supply chains, from mining to recycling, and reuse.

In sum, DSMS provides user-friendly solutions for all actors in a product life cycle, including experimenters, metallurgy experts, modelling experts, data scientists, product designers (even with little or no programming background). It not only makes data handling more efficient and increases productivity, but also enhances collaboration among diverse stakeholders. This contrasts with traditional systems, were typically only few contact points between the actors exist, hindering mutual understanding and the ability to drive innovations. In the future, DSMS will further establish its role as an enabler, bridging existing systems and AI-driven tools, meaning a more prominent presence of AI both in the user experience and behind the scenes. Moreover, a robust app store, process design tools, and improved control over connected external data sources will be further developed in the near future. On the application side, the next steps involve identify additional use cases from both industrial and scientific domains, with the goal of fully showcasing the potential of semantic data management through DSMS. By doing so, DSMS will continue to evolve in terms of both functionality and user experience. The practical relevance of the DSMS-powered dataspace solutions showcased in the StahlDigital and KupferDigital projects highlights the transformative potential of semantic technologies in materials science and manufacturing.

**Author contribution**

YN wrote the manuscript together with LM, DH, and MW. Furthermore, YN orchestrated and contributed to the development of DSMS. LM worked on validating DSMS in the context of research project StahlDigital and on the presented KupferDigital use case. MB contributed to the development of the DSMS backend and developed the DSMS-SDK as well as the new version of data2rdf. He also worked on the semantic material card use case of StahlDigital. MW supported regarding the presented KupferDigital use case. DH is the principal investigator for the StahlDigital Fraunhofer IWM part and orchestrate the project including the strategic direction and user requirements. PZ developed the first version of data2rdf, contributed to the development of DSMS and validated it in the context of research project StahlDigital. KK developed the first version of form2rdf and the visualization functionalities of DSMS. PA contributed to the vocabulary service of DSMS as well as supporting work in its backend and frontend. All authors contributed to the discussion of the results.

## Supporting information

Data from project StahlDigital can be accessed via the StahlDigital DSMS instance at https://stahldigital.materials-data.space or upon reasonable request. Data from project KupferDigital can be accessed via the KupferDigital DSMS instance at https://kupferdigital.materials-data.space or at https://www.zenodo.org/records/10820299 [75]. For data integration, we used the ontology developed in project StahlDigital [67]. For data integration we used the tool data2rdf [43]. For interacting with DSMS (besides using the frontend) and for programming apps, we used the Python library DSMS-SDK [45].

## Acknowledgements

The authors thankfully acknowledge support by the German Federal Ministry of Education and Research (BMBF) through the funding initiative MaterialDigital for the projects StahlDigital and KupferDigital, with grant numbers 13XP5116 and 13XP5119 respectively.

Received: ((will be filled in by the editorial staff))
Revised: ((will be filled in by the editorial staff))
Published online: ((will be filled in by the editorial staff))

## References

1. Panchal JH, Kalidindi SR, McDowell DL (2013) Key computational modeling issues in Integrated Computational Materials Engineering. Computer-Aided Design 45:4–25. https://doi.org/10.1016/j.cad.2012.06.006

2. Shrouf F, Ordieres J, Miragliotta G Smart factories in Industry 4.0: A review of the concept and of energy management approached in production based on the Internet of Things paradigm

3. Wilkinson MD, Dumontier M, Aalbersberg IJJ et al. (2016) The FAIR Guiding Principles for scientific data management and stewardship. Scientific data 3:160018. https://doi.org/10.1038/sdata.2016.18

4. Curry E (2020) Dataspaces: fundamentals, principles, and techniques. Real-time Linked Dataspaces: Enabling Data Ecosystems for Intelligent Systems:45–62

5. Hogan A, Blomqvist E, Cochez M et al. (2022) Knowledge Graphs. ACM Comput Surv 54:1–37. https://doi.org/10.1145/3447772

6. Agrawal A, Choudhary A (2016) Perspective: Materials informatics and big data: Realization of the "fourth paradigm" of science in materials science. APL Mater 4:53208. https://doi.org/10.1063/1.4946894

7. Dornheim J, Morand L, Nallani HJ et al. (2024) Neural Networks for Constitutive Modeling: From Universal Function Approximators to Advanced Models and the Integration of Physics. Arch Computat Methods Eng 31:1097–1127. https://doi.org/10.1007/s11831-023-10009-y

8. Dornheim J, Morand L, Zeitvogel S et al. (2022) Deep reinforcement learning methods for structure-guided processing path optimization. J Intell Manuf 33:333–352. https://doi.org/10.1007/s10845-021-01805-z

9. Iraki T, Morand L, Dornheim J et al. (2023) A multi-task learning-based optimization approach for finding diverse sets of microstructures with desired properties. J Intell Manuf. https://doi.org/10.1007/s10845-023-02139-8

10. Morand L, Norouzi E, Weber M et al. (2024) Data-Driven Accelerated Parameter Identification for Chaboche-Type Visco-Plastic Material Models to Describe the Relaxation Behavior of Copper Alloys. Exp Mech. https://doi.org/10.1007/s11340-024-01057-x

11. Durmaz AR, Müller M, Lei B et al. (2021) A deep learning approach for complex microstructure inference. Nature communications 12:6272. https://doi.org/10.1038/s41467-021-26565-5

12. Kalidindi SR, Buzzy M, Boyce BL et al. (2022) Digital Twins for Materials. Front Mater 9. https://doi.org/10.3389/fmats.2022.818535

13. Morand L, Link N, Iraki T et al. (2022) Efficient Exploration of Microstructure-Property Spaces via Active Learning. Front Mater 8. https://doi.org/10.3389/fmats.2021.824441

14. Shoghi R, Morand L, Helm D et al. (2024) Optimizing machine learning yield functions using query-by-committee for support vector classification with a dynamic stopping criterion. Comput Mech. https://doi.org/10.1007/s00466-023-02440-6

15. Morand L, Helm D (2019) A mixture of experts approach to handle ambiguities in parameter identification problems in material modeling. Computational Materials Science 167:85–91. https://doi.org/10.1016/j.commatsci.2019.04.003

16. Stevens R, Goble CA, Bechhofer S (2000) Ontology-based knowledge representation for bioinformatics. Briefings in bioinformatics 1:398–414. https://doi.org/10.1093/bib/1.4.398

17. Navigli R, Ponzetto SP (2012) BabelNet: The automatic construction, evaluation and application of a wide-coverage multilingual semantic network. Artificial Intelligence 193:217–250. https://doi.org/10.1016/j.artint.2012.07.001

18. Vrandečić D, Krötzsch M (2014) Wikidata. Commun ACM 57:78–85. https://doi.org/10.1145/2629489
19. Lehmann J, Isele R, Jakob M et al. (2015) DBpedia – A large-scale, multilingual knowledge base extracted from Wikipedia. SW 6:167–195. https://doi.org/10.3233/SW-140134
20. Noy N, Gao Y, Jain A et al. (2019) Industry-scale knowledge graphs. Commun ACM 62:36–43. https://doi.org/10.1145/3331166
21. Krishnan A (2018) Making search easier. https://www.aboutamazon.com/news/innovation-at-amazon/making-search-easier
22. Singhal A (2012) Introducing the Knowledge Graph: things, not strings. https://blog.google/products/search/introducing-knowledge-graph-things-not/
23. Shrivastava S (2017) Bring rich knowledge of people, places, things and local businesses to your apps. https://blogs.bing.com/search-quality-insights/2017-07/bring-rich-knowledge-of-people-places-things-and-local-businesses-to-your-apps
24. Chang S (2018) Scaling Knowledge Access and Retrieval at Airbnb. https://medium.com/airbnb-engineering/scaling-knowledge-access-and-retrieval-at-airbnb-665b6ba21e95
25. Bayerlein B, Hanke T, Muth T et al. (2022) A perspective on digital knowledge representation in materials science and engineering. Advanced Engineering Materials 24:2101176
26. Francisco Morgado J, Chiacchiera S, Le Franc Y et al. (2023) A report on the Workshop "Towards Materials and Manufacturing Commons - the enablers Digital Marketplaces, FAIR Principles and Ontologies"
27. EMMC - The European Materials Modelling Council EMMC Roadmap 2023: Digital Transformation of Materials Science
28. Bechhofer S, Buchan I, Roure D et al. (2013) Why linked data is not enough for scientists. Future Generation Computer Systems 29:599–611. https://doi.org/10.1016/j.future.2011.08.004
29. EMMC - The European Materials Modelling Council (2024) Elementary Multiperspective Material Ontology (EMMO)
30. Arp R, Smith B, Spear AD (2015) Building Ontologies with Basic Formal Ontology. The MIT Press

31. Bayerlein B, Schilling M, Birkholz H et al. (2024) PMD Core Ontology: Achieving semantic interoperability in materials science. Materials & Design 237:112603. https://doi.org/10.1016/j.matdes.2023.112603

32. Codd EF (1970) A relational model of data for large shared data banks. Commun ACM 13:377–387. https://doi.org/10.1145/362384.362685

33. Higgins SG, Nogiwa-Valdez AA, Stevens MM (2022) Considerations for implementing electronic laboratory notebooks in an academic research environment. Nature protocols 17:179–189. https://doi.org/10.1038/s41596-021-00645-8

34. Himanen L, Geurts A, Foster AS et al. (2019) Data-Driven Materials Science: Status, Challenges, and Perspectives. Advanced science (Weinheim, Baden-Wurttemberg, Germany) 6:1900808. https://doi.org/10.1002/advs.201900808

35. Rowley J (2007) The wisdom hierarchy: representations of the DIKW hierarchy. Journal of Information Science 33:163–180. https://doi.org/10.1177/0165551506070706

36. International Data Spaces Association (IDSA). https://internationaldataspaces.org/

37. Roters F, Aslam A, Bai Y et al. StahlDigital: Ontology-based workflows for the steel industry. Advanced Engineering Materials (to be published) 2024

38. Eisenbart M, Hanke T, Bauer F et al. (2024) KupferDigital towards a circular economy: Ontology-based digital representation for the copper life-cycle. Advanced Engineering Materials (to be published)

39. (2022) IEEE/ISO/IEC International Standard for Software, systems and enterprise-- Architecture description. ISO/IEC/IEEE 42010:2022(E). https://doi.org/10.1109/IEEESTD.2022.9938446

40. Hogan A (2020) The Semantic Web: Two decades on. SW 11:169–185. https://doi.org/10.3233/SW-190387

41. Suárez-Figueroa MC, Gómez-Pérez A, Fernández-López M (2015) The NeOn Methodology framework: A scenario-based methodology for ontology development. Applied Ontology 10:107–145. https://doi.org/10.3233/AO-150145

42. Poveda-Villalón M, Fernández-Izquierdo A, Fernández-López M et al. (2022) LOT: An industrial oriented ontology engineering framework. Engineering Applications of Artificial Intelligence 111:104755. https://doi.org/10.1016/j.engappai.2022.104755

43. Zierep. Paul, Büschelberger M, Nahshon Y et al. data2rdf. https://github.com/MI-FraunhoferIWM/data2rdf

44. Reimers N, Gurevych I Sentence-BERT: Sentence Embeddings using Siamese BERT-Networks
45. Büschelberger M dsms-python-sdk. https://github.com/MI-FraunhoferIWM/dsms-python-sdk
46. Charles R. Harris, K. Jarrod Millman, Stéfan J. van der Walt et al. (2020) Array programming with NumPy. Nature 585:357–362. https://doi.org/10.1038/s41586-020-2649-2
47. Stéfan van der Walt, Jarrod Millman (eds) (2010) Proceedings of the 9th Python in Science Conference
48. Fabian Pedregosa, Gaël Varoquaux, Alexandre Gramfort et al. (2011) Scikit-learn: Machine Learning in Python. Journal of Machine Learning Research 12:2825–2830
49. Martín Abadi, Ashish Agarwal, Paul Barham et al. (2016) TensorFlow: Large-Scale Machine Learning on Heterogeneous Distributed Systems
50. Adam Paszke, Sam Gross, Francisco Massa et al. (2019) PyTorch: An Imperative Style, High-Performance Deep Learning Library
51. World Wide Web Consortium, others (2014) Data catalog vocabulary (DCAT)
52. Weibel S, Kunze J, Lagoze C et al. (1998) Dublin core metadata for resource discovery
53. QUDT board QUDT ontology. https://www.qudt.org/
54. Goldbeck G, Simperler A, Andres P de et al. (2023) MarketPlace - a Digital Materials Modelling Marketplace
55. Materials Modelling Marketplace for Increased Industrial Innovation. https://cordis.europa.eu/project/id/760173
56. Virtual Materials Market Place (VIMMP). https://cordis.europa.eu/project/id/760907
57. Digital Open Marketplace Ecosystem 4.0. https://cordis.europa.eu/project/id/953163
58. Argo Workflows. https://argoproj.github.io/
59. Markus Scheidgen, Lauri Himanen, Alvin Noe Ladines et al. (2023) NOMAD: A distributed web-based platform for managing materials science research data. Journal of Open Source Software 8:5388. https://doi.org/10.21105/joss.05388
60. European Organization For Nuclear Research, OpenAIRE (2013) Zenodo. https://www.zenodo.org/
61. figshare. https://figshare.com/
62. CKAN. https://www.ckan.org/
63. Siska V, Karagiannis V, Drobics M et al. (2023) Building a Dataspace: Technical Overview. Gaia-X Hub Austria

64. Michael Neubauer, Lukas Steinle, Colin Reiff et al. (2023) Architecture for manufacturing-X: Bringing asset administration shell, eclipse dataspace connector and OPC UA together. Manufacturing Letters 37:1–6. https://doi.org/10.1016/j.mfglet.2023.05.002

65. Keycloak Community Keycloak. https://www.keycloak.org

66. Thorgersen S, Silva PI (2021) Keycloak-identity and access management for modern applications: harness the power of Keycloak, OpenID Connect, and OAuth 2.0 protocols to secure applications. Packt Publishing Ltd

67. Meyer L-P, et al. SteelProcessOntology. https://w3id.org/steel/ProcessOntology

68. Lebo T, Sahoo S, McGuinness D et al. (2013) Prov-o: The prov ontology. W3C recommendation 30

69. Eisenbart M, Friedmann V, Preußner J et al. (2021) Entwicklung und Charakterisierung von ausscheidungshärtenden Legierungen im System Cu-Ni-Al. https://publica.fraunhofer.de/handle/publica/420094

70. Hossein Beygi Nasrabadi, Felix Bauer, Ladji Tikana et al. (2024) KupferDigital mechanical testing datasets: Brinell hardness, Vickers hardness, and tensile tests

71. Schilling M, Bayerlein B, Hartrott P von et al. (2024) FAIR and Structured Data: A Domain Ontology Aligned with Standard-Compliant Tensile Testing. Advanced Engineering Materials:2400138

72. Klein P, Konchakova N, Hristova-Bogaerds DG et al. (2021) Translation in materials modelling: Process and progress. OntoTrans-FORCE, White Paper

73. Braud A, Fromentoux G, Radier B et al. (2021) The road to European digital sovereignty with Gaia-X and IDSA. IEEE network 35:4–5

74. Ehlers L (2023) The impact of Catena-X on digital business models in the automotive industry

75. Nasrabadi HB, Bauer F, Tikana L et al. KupferDigital mechanical testing datasets: Brinell hardness, Vickers hardness, and tensile tests. https://zenodo.org/records/10820299